\documentclass[5p,superscriptaddress]{revtex4-2}

\usepackage{xcolor}
\usepackage{graphicx} 
\usepackage{subfigure}
\usepackage{dcolumn} 
\usepackage{bm}
\usepackage{float}
\usepackage{hyperref}
\usepackage{amsmath}
\usepackage{caption}
\usepackage{textgreek}

\usepackage{tikz,xcolor,hyperref}

\definecolor{lime}{HTML}{A6CE39}
\DeclareRobustCommand{\orcidicon}{
	\begin{tikzpicture}
	\draw[lime, fill=lime] (0,0) 
	circle [radius=0.16] 
	node[white] {{\fontfamily{qag}\selectfont \tiny ID}};
	\draw[white, fill=white] (-0.0625,0.095) 
	circle [radius=0.007];
	\end{tikzpicture}
	\hspace{-2mm}
}
\foreach \x in {A, ..., Z}{%
	\expandafter\xdef\csname orcid\x\endcsname{\noexpand\href{https://orcid.org/\csname orcidauthor\x\endcsname}{\noexpand\orcidicon}}
}
\foreach \x in {A, ..., Z}{%
	\expandafter\xdef\csname orcid\x\endcsname{\noexpand\href{https://orcid.org/\csname orcidauthor\x\endcsname}{\noexpand\orcidicon}}
}

\begin{document}

\title{A Review on Intense Electromagnetic Fields in Heavy-Ion Collisions: Theoretical Predictions and Experimental Results}

\author{Diyu Shen\orcidA{}}
\affiliation{Key Laboratory of Nuclear Physics and Ion-beam Application (MOE), Institute of Modern Physics, Fudan University, Shanghai, 200433, China}
\affiliation{Heavy Ion Science and Technology Key Laboratory, Institute of Modern Physics, Chinese Academy of Sciences, Lanzhou 730000, China}

\author{Jinhui Chen\orcidB{}}
\affiliation{Key Laboratory of Nuclear Physics and Ion-beam Application (MOE), Institute of Modern Physics, Fudan University, Shanghai, 200433, China}
\affiliation{Shanghai Research Center for Theoretical Nuclear Physics, NSFC and Fudan University, Shanghai 200438, China}

\author{Xu-Guang Huang\orcidH{}}
\affiliation{Physics Department and Center for Particle Physics and Field Theory, Fudan University, Shanghai, 200438, China}
\affiliation{Key Laboratory of Nuclear Physics and Ion-beam Application (MOE), Institute of Modern Physics, Fudan University, Shanghai, 200433, China}
\affiliation{Shanghai Research Center for Theoretical Nuclear Physics, NSFC and Fudan University, Shanghai 200438, China}

\author{Yu-Gang Ma\orcidD{}}
\email{mayugang@fudan.edu.cn}
\affiliation{Key Laboratory of Nuclear Physics and Ion-beam Application (MOE), Institute of Modern Physics, Fudan University, Shanghai, 200433, China}
\affiliation{Shanghai Research Center for Theoretical Nuclear Physics, NSFC and Fudan University, Shanghai 200438, China}

\author{Aihong Tang\orcidT{}}
\affiliation{Brookhaven National Laboratory, Upton, New York 11973, USA}

\author{Gang Wang\orcidW{}}
\affiliation{Department of Physics and Astronomy, University of California, Los Angeles, California 90095, USA}

\date{\today}

\begin{abstract}
 In heavy-ion collisions at relativistic energies, the incident nuclei travel at nearly the speed of light. These collisions deposit kinetic energy into the overlap region and create a high-temperature environment where hadrons ``melt'' into deconfined quarks and gluons. The spectator nucleons, which do not undergo scatterings, generate an ultra-intense electromagnetic field---on the order of $10^{18}$ Gauss at Relativistic Heavy-Ion Collider, and $10^{19}$ Gauss at the Large Hadron Collider. These powerful electromagnetic fields have a significant impact on the produced particles, not only complicating the study of particle interactions but also inducing novel physical phenomena. To explore the nature of these fields and their interactions with deconfined quarks, we provide a detailed overview, encompassing theoretical estimations of their generation and evolution, as well as experimental efforts to detect them. We also provide physical interpretations of the discovered results and discuss potential directions for future investigations. 
\end{abstract}

\maketitle

\section{Introduction}
Heavy-ion collision experiments investigate the properties of nuclear matter under extreme temperatures and energy densities by smashing atomic nuclei at ultra-high center-of-mass energies ($\sqrt{s_{NN}}$). A primary goal is to create and study the quark-gluon plasma (QGP)~\cite{Shuryak:1980tp}, a state of matter in which quarks and gluons are no longer confined within color-neutral hadrons, resembling conditions in the early universe following the Big Bang~\cite{BigBang-Kolb}. The two major facilities conducting such experiments are Relativistic Heavy-Ion Collider (RHIC) at Brookhaven National Laboratory (BNL) and the Large Hadron Collider (LHC) at CERN. At these laboratories, heavy ions, such as copper (Cu), gold (Au), and lead (Pb), are accelerated to relativistic speeds (greater than 0.99$c$, where $c$ denotes the speed of light in vacuum). Measurements from such experiments provide crucial insights into quantum chromodynamics (QCD), including the phase transition between ordinary nuclear matter and QGP, chiral symmetry restoration, and the violation of ${\cal P}$ (parity) and ${\cal CP}$ (charge-parity) symmetries in strong interactions~\cite{Ko_NST,Chen,Shou,Liu:2020ymh}.

In heavy-ion collisions, the relativistic charged protons within nuclei generate intense magnetic fields, reaching strengths of approximately 
$10^{18}$ Gauss at RHIC and $10^{19}$ Gauss at the LHC~\cite{Skokov:2009qp,Kharzeev:2007jp,Deng:2012pc}. Such immense field strengths make electromagnetic interactions non-negligible compared with the energy scale of the strong interaction in the QGP~\cite{Yan:2021zjc}.
Initial interest in these magnetic fields arose from studies of the chiral magnetic effect (CME) in heavy-ion collisions~\cite{Kharzeev:2007jp,Fukushima:2008xe}. 
The CME describes the induction of 
an electric current along the magnetic field direction if the quarks in the QGP possess a net chirality~\cite{Kharzeev:2007jp,Fukushima:2008xe,Kharzeev:2015znc,Kharzeev:2024zzm}.
Chirality refers to the intrinsic handedness of quarks: right-handed (left-handed) quarks have spin projections aligned (anti-aligned) with their momentum direction, whereas for anti-particles, this relationship is reversed.
In addition to the deconfinement phase transition, light quarks are expected to undergo a chiral phase transition at high temperatures, restoring chiral symmetry, becoming almost massless, and acquiring a definite chirality~\cite{Kharzeev:2020jxw}.
Meanwhile, gluon fields in the QCD vacuum can undergo topological transitions that violate ${\cal P}$ and ${\cal CP}$ symmetries, transferring a net chirality to quarks in the QGP~\cite{Kharzeev:2007jp,Fukushima:2008xe,Kharzeev:2015znc}.
Thus, heavy-ion collisions can satisfy both necessary conditions for the CME: a strong magnetic field and a net chirality in the QGP.
In addition to the CME, strong magnetic fields can significantly influence QCD phase transitions. They may act as catalysts for chiral symmetry breaking at low temperature~\cite{Gusynin:1994re}, suppress the temperatures at which chiral phase transitions occur~\cite{Bali:2011qj,Bali:2012zg}, and alter the critical endpoint in the QCD phase diagram, where the phase transition shifts from first-order to a crossover ~\cite{Mizher:2010zb,Fraga:2008qn,Endrodi:2015oba}.
Strong magnetic fields are also prevalent in astrophysical systems, such as neutron stars, where surface magnetic fields can reach 
$10^{12}$--$10^{13}$ Gauss. These extreme conditions have spurred research into neutron star structure and the equation of state of neutron matter under intense magnetic fields~\cite{Bigdeli:2017uzy,Yakhshiev:2019gvb,Andreichikov:2013pga}. Heavy-ion collision experiments offer a unique opportunity to investigate these phenomena in a controlled laboratory setting. Because of these interests, this article mainly focuses on the magnetic field, although the electric field could also be non-negligible on an event-by-event basis. 

Although classical electrodynamics can reliably estimate the maximum strength of the magnetic field in heavy-ion collisions~\cite{Skokov:2009qp,Kharzeev:2007jp,Deng:2012pc,Zhao:2019crj,Zhao}, its temporal evolution remains an open theoretical question. The complexities arise from the interplay between the magnetic field and the charge-conducting medium~\cite{Tuchin:2015oka,Huang:2022qdn,Li:2023tsf,Wang:2021oqq,Yan:2021zjc,Zhang:2022lje,Huang:2024aob}, which is governed by Maxwell's equations and highly sensitive to the temperature-dependent electric conductivity of the medium~\cite{Ding:2010ga,Francis:2011bt,Amato:2013naa}. The pre-equilibrium properties of the medium also shape the magnetic field's strength, as the system transitions into hydrodynamic evolution~\cite{Yan:2021zjc}. Additionally, deviations from Ohm's law in the medium may affect the field's evolution~\cite{Wang:2021oqq}, while recent studies suggest that the QGP fluid vorticity can extend its lifetime~\cite{Guo:2019mgh,Huang:2024aob}. Due to these complexities, estimates of the magnetic field strength in the medium can vary by several orders of magnitude across different models~\cite{McLerran:2013hla,Wang:2021oqq,Li:2023tsf,Yan:2021zjc,Voronyuk:2011jd,Huang:2024aob,Stewart:2021mjz}. 

The significant theoretical uncertainties highlight the need for experimental investigations of the magnetic field in heavy-ion collisions. Various phenomena induced by a strong magnetic field can serve as experimental magnetometers, including the CME~\cite{Kharzeev:2007jp,Kharzeev:2020jxw,Fukushima:2008xe,Kharzeev:2015znc}, charge-dependent collective motion~\cite{Mohapatra:2011ku,Tuchin:2011jw,Das:2016cwd,Dubla:2020bdz,Gursoy:2018yai,Gursoy:2014aka,Nakamura:2022ssn,Voronyuk:2014rna}, the polarization difference between $\Lambda$ and $\bar{\Lambda}$~\cite{Peng:2022cya}, transverse momentum ($p_T$) spectra of dilepton pairs~\cite{STAR:2019wlg,Brandenburg:2021lnj}, directed flow of direct photons~\cite{Sun:2023rhh,ALICE:2018dti,PHENIX:2015igl}, and baryon electric-charge correlations~\cite{Ding:2023bft,Huang:NST}. In this article, we review and discuss recent experimental results related to these effects.

We organize this paper as follows. In Section \ref{sec:theory}, we discuss theoretical predictions of the magnetic field, including its strength and temporal evolution across different models. Section \ref{sec:exp} reviews and analyzes experimental findings from various observables related to magnetic field detection. In Section \ref{sec:cha}, we outline key challenges and propose directions for future research. Finally, we summarize our conclusions in Section \ref{sec:sum}.

\section{The Electromagnetic Field Production}\label{sec:theory}
\subsection{Nuclear-spectator-induced electromagnetic field}
\begin{figure}[htbp]
\vspace*{-0.1in}
\includegraphics[scale=0.3]{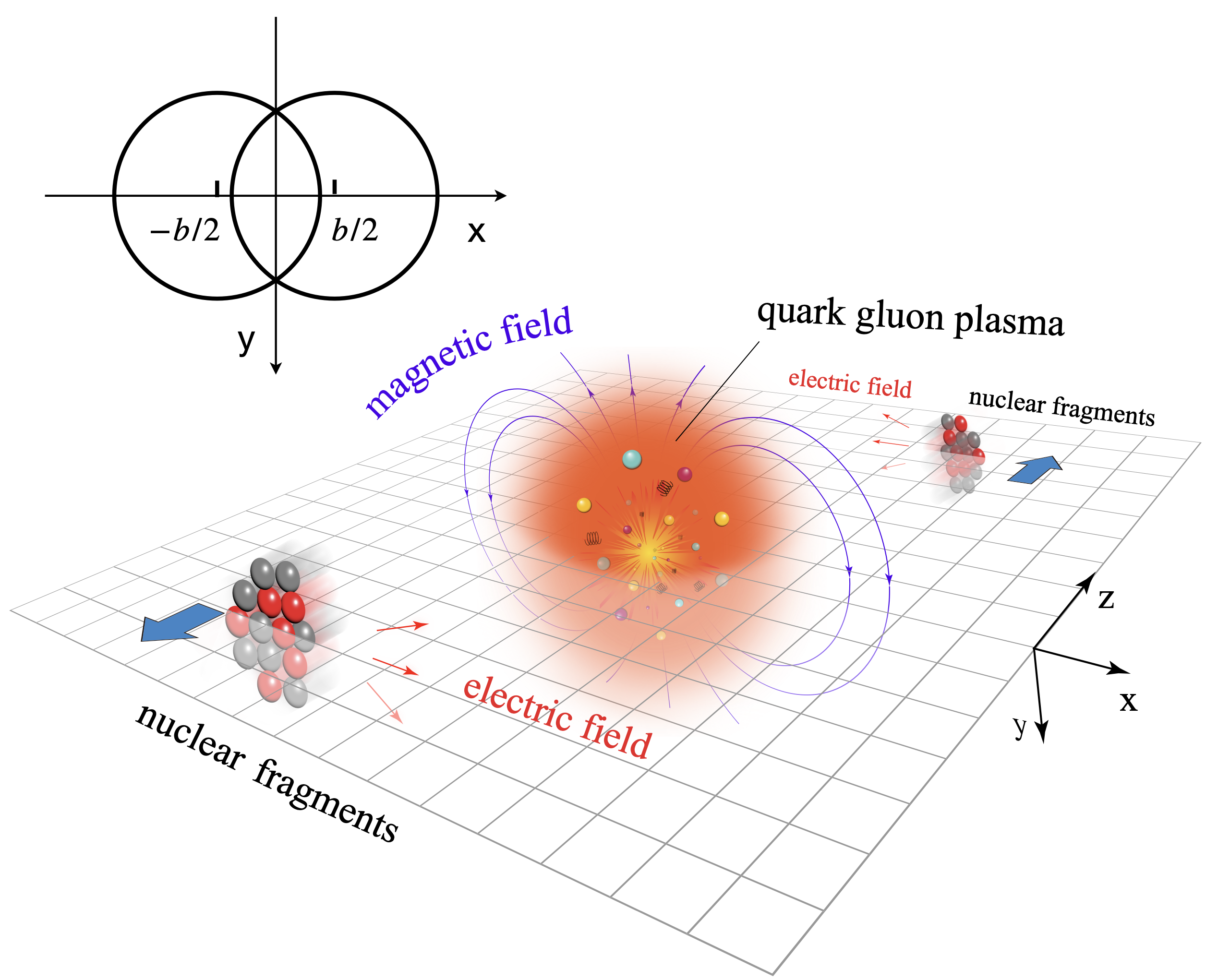}
\captionof{figure}{Sketch of a heavy-ion collision in the lab frame (figure is from Ref.~\cite{STAR:2023jdd}).
The impact parameter and the beam direction are along the $x$ and $z$ axes, respectively. 
Spectator nuclear fragments generate strong magnetic fields along $-y$. }
\label{fig:illustration}
\end{figure}

In heavy-ion collision experiments at relativistic energies, positively charged nuclei are accelerated to velocities approaching the speed of light. When these nuclei collide, the overlapping region can experience significant energy deposition, potentially generating a deconfined QGP. However, in most collisions, the nuclei do not fully overlap. The transverse distance between the centers of the two incident nuclei is defined as the impact parameter $b$, as illustrated in Fig.~\ref{fig:illustration}. Collisions with $b=0$ are conceptually referred to as the most central collisions, while those with $b$ greater than the sum of the radii of the two nuclei are classified as ultra-peripheral collisions (UPC). In collisions with finite $b$, spectator nucleons remain outside the overlapping region and do not directly participate in the interaction. The spectator protons, carrying positive charges and moving rapidly along the beam direction, generate an intense magnetic field in their vicinity.

\begin{widetext}
\begin{figure*}[htbp]
\vspace*{-0.1in}
\includegraphics[scale=0.35]{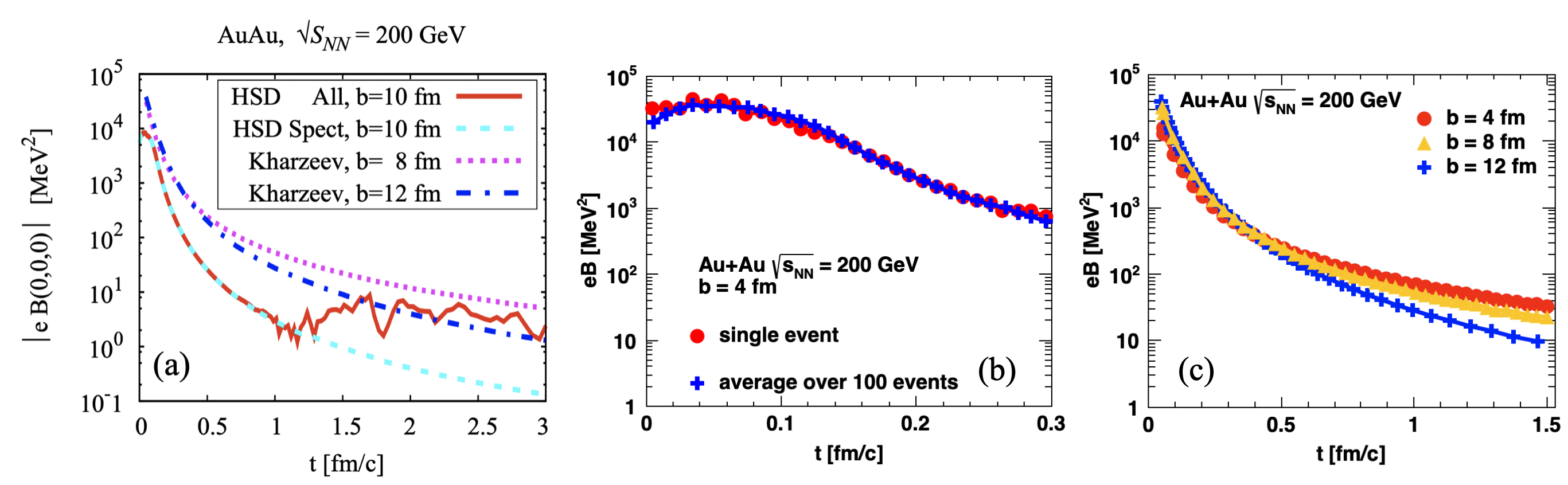}
\caption{Magnetic field produced mainly by spectator protons in semi-central Au+Au collisions at $\sqrt{s_{NN}}=$200 GeV, without medium responses (figures are from Refs.~\cite{Voronyuk:2011jd,Skokov:2009qp,Kharzeev:2007jp}). }
\label{fig:peakMag}
\end{figure*}
\end{widetext}

The strength of the produced magnetic field can be calculated using the Liénard-Wiechert potentials, which are solutions to Maxwell's equations that describe the electromagnetic field produced by a moving point charge. Based on these potentials, the magnetic and electric fields are given by the following expressions:
\begin{eqnarray}
e{\bf B}(t,{\bf r})\ =\ \frac{e^{2}}{4\pi}\sum_{n}Z_{n}\frac{{\bf v}_{n}\times{\bf R}_{n}}{(R_{n}-{\bf R}_{n}\cdot{\bf v}_{n})^{3}}(1-|{\bf v}_{n}|^{2}), \label{Eq:LW_B}\\
e{\bf E}(t,{\bf r})\ =\ \frac{e^{2}}{4\pi}\sum_{n}Z_{n}\frac{{\bf R}_{n}-R_{n}{\bf v}_{n}}{(R_{n}-{\bf R}_{n}\cdot{\bf v}_{n})^{3}}(1-|{\bf v}_{n}|^{2}),
\label{Eq:LW_E}
\end{eqnarray}
where $Z_{n}$ is the electric charge of the $n{\rm th}$ particle, ${\bf R}_{n}= {\bf r} - {\bf r}_n$ is the distance vector between the field point $\bf r$ and the position ${\bf r}_n$ of the $n{\rm th}$ particle with velocity ${\bf v}_{n}$ at the retarded time $t_n = t-|{\bf r}-{\bf r}_n|$. $t$ represents the moment at which the electromagnetic field is measured. These equations are fundamental in electrodynamics and are widely used to calculate the electric and magnetic fields generated by a charged particle moving with an arbitrary velocity. Although the Liénard-Wiechert potential is a classical formulation, it effectively describes the generation of electromagnetic fields in heavy-ion collisions, as quantum corrections to the Liénard-Wiechert potential play only a minor role in this context~\cite{Huang:2015oca}. The primary uncertainty in calculating the produced electromagnetic field arises from the uncertain spatial distribution of spectator protons. An empirical model commonly used in nuclear physics, the Woods-Saxon distribution, describes the radial distribution of nuclear matter within an atomic nucleus, particularly for the nuclear potential or the spatial distribution of nucleons inside a nucleus. It is given by the following form~\cite{Shou_PLB,STAR:2024wgy,Giacalone_NST,Xu:2024bdh,Wang:2024vjf,Schenke_NST}:
\begin{eqnarray}
\rho(r,\theta,\phi)&=&\frac{\rho_{0}}{1+e^{\left[\frac{r-R(\theta)}{a}\right]}}, \\
R(\theta)&=&R_{0}\left[1+\beta_{2}Y_{2}^{0}(\theta)+\beta_{4}Y_{4}^{0}(\theta)+\ldots.\right],
\end{eqnarray}
where $R_{0}$ is the spherical radius, $\beta_2$ and $\beta_4$ are the deformation parameters (with $\beta_2$ typically representing the quadrupole deformation), and $Y_{2}^{0}(\theta)$ and $Y_{4}^{0}(\theta)$ are the spherical harmonics. The deformation parameters should be determined from experiments. For nearly spherical nuclei, such as gold (Au), a symmetric nucleon distribution is often used to estimate the strength of the electromagnetic field in Au+Au collisions at $\sqrt{s_{NN}}=$ 200 GeV per nucleon pair, as shown in Fig.~\ref{fig:peakMag}. The peak strength of the magnetic field in heavy-ion collisions at RHIC could reach $\sim 10^{18}$ Gauss (1 Gauss $\approx 1.95\times10^{-20}$ $\rm GeV^2$), which is ten thousand times greater than the magnetic field of magnetars~\cite{Kouveliotou:1998ze}. However, such a magnetic field decays quickly as the spectator protons move away. Figure~\ref{fig:peakMag} shows the time evolution of the magnetic field in a vacuum, where it becomes negligible compared with the peak value in $\sim 10^{-24}$ second (1 fm/$c$ $\approx$ $3.3 \times 10^{-24}$ s) after the collision. Nevertheless, the presence of the QGP could prolong the lifetime of the magnetic field due to the Faraday induction effect, which will be discussed in the next section. 

\begin{widetext}
\begin{figure*}[htbp]
\vspace*{-0.1in}
\includegraphics[scale=0.25]{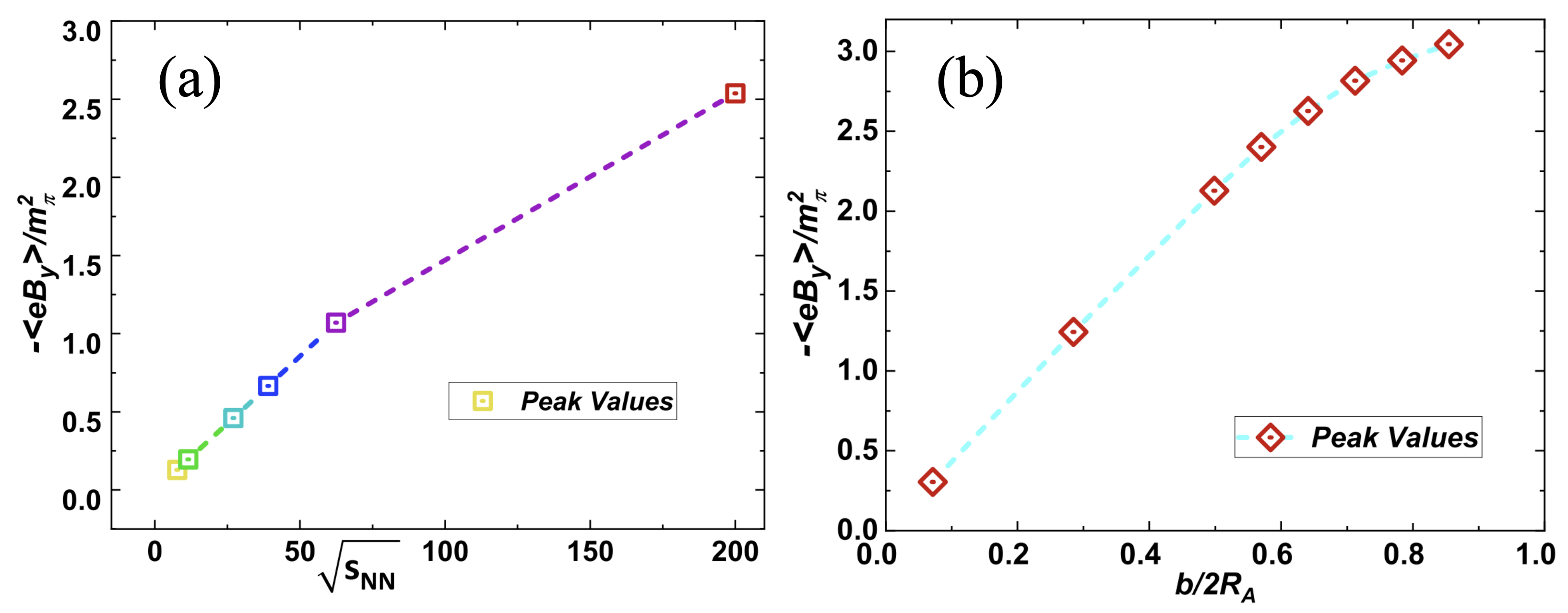}
\captionof{figure}{The peak value of the magnetic field strength as a function of (a) collision energy and (b) impact parameter (figure is from Ref.~\cite{Siddique:2021smf}). }
\label{fig:peakBVsBeam}
\end{figure*}
\end{widetext}

The Liénard-Wiechert potentials show that the magnetic field generated by a moving charge depends on both the velocity and the number of source charges. This implies that the peak magnetic field strength is larger at higher collision energies and for larger impact parameters $b$ as long as $b\lesssim 2 R_0$. Such a relationship has been confirmed in the Refs.~\cite{Skokov:2009qp,Deng:2012pc,Siddique:2021smf} as shown in Fig.~\ref{fig:peakBVsBeam}. 
On the other hand, the initial structure of the colliding nuclei \cite{MaYG2023,Jia_NST} is also important for the distribution of 
electromagnetic fields. For example, Ref.~\cite{ChengYL_PRC} observed that the electric and magnetic fields display different behavioral patterns for asymmetric combinations of the projectile and target nuclei as well as for different initial configurations of the nucleus.

\subsection{Event-by-event magnetic field fluctuations}
\begin{widetext}
\begin{figure*}[htbp]
\vspace*{-0.1in}
\includegraphics[scale=0.3]{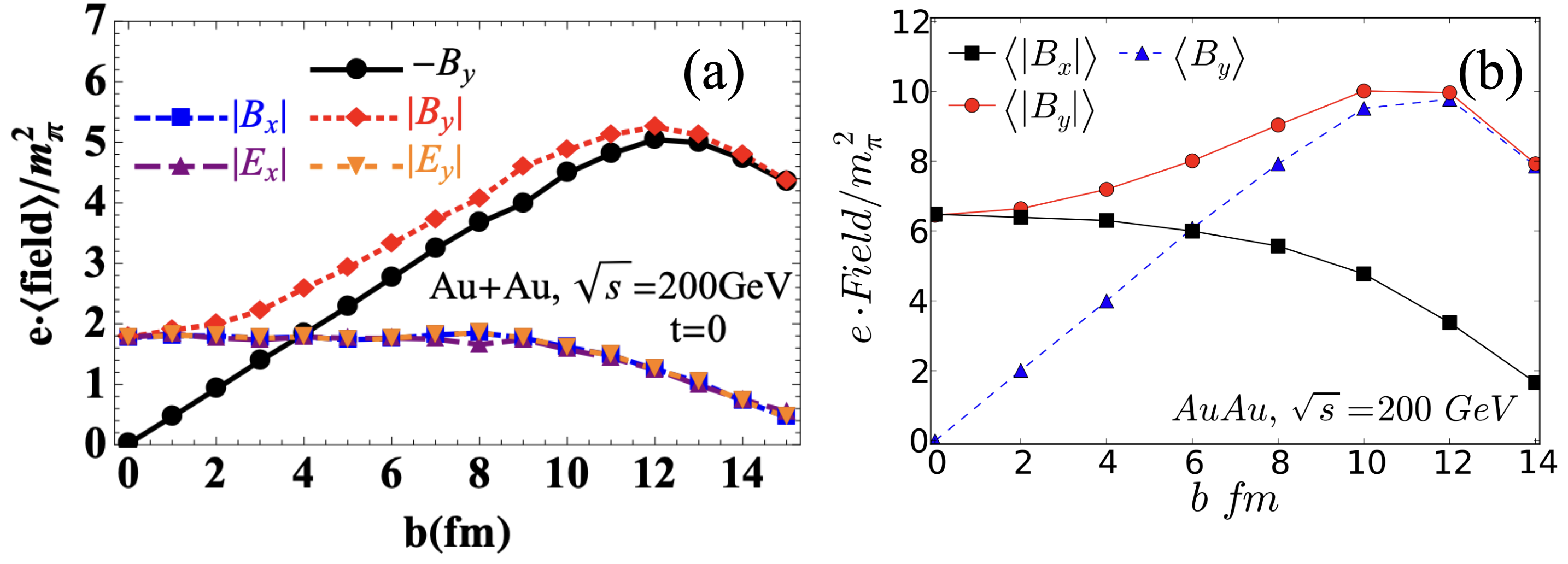}
\captionof{figure}{Electromagnetic fields in Au+Au collisions at $\sqrt{s_{NN}}=$ 200 GeV as functions of impact parameter $b$. The magnetic- and electric-field fluctuations are indicated by the magnitude averages along different directions. (a) Calculations using initial condition from the Heavy Ion Jet INteraction Generator (HIJING) model (figure is from Ref.~\cite{Deng:2012pc}). (b) The initial nucleon distribution is from the Woods-Saxon distribution with standard parameters, assuming both nuclei are infinitely thin (figure is from Ref.~\cite{Bzdak:2011yy}). }
\label{fig:EMFluctuation}
\end{figure*}
\end{widetext}

The spectator-induced magnetic field is, on average, aligned along the $-y$ axis over multiple events, as illustrated in Fig.~\ref{fig:illustration}, due to the mirror symmetry of the collision geometry. However, on an event-by-event basis, it fluctuates significantly~\cite{Deng:2012pc,Bzdak:2011yy,Bloczynski:2012en,Bloczynski:2013mca,Zhao:2019crj}. The strength and direction of the magnetic field depend on the spatial distribution and velocity of spectator protons, as indicated in Eq.~\ref{Eq:LW_B}. Although the velocity of protons is determined by the collision energy and beam direction—both known parameters—the spatial distribution of nucleons fluctuates event by event due to their quantum nature, leading to fluctuations in the magnetic field. Notably, it has been found that the magnetic field component along the impact parameter direction, $B_x$, can be comparable to $B_y$, as demonstrated in Fig.~\ref{fig:EMFluctuation}~\cite{Deng:2012pc,Bzdak:2011yy}. Additionally, the electric field components, $E_x$ and $E_y$, can be quite strong and are approximately equal in magnitude. More quantitatively, results from Ref.~\cite{Deng:2012pc} in Fig.~\ref{fig:EMFluctuation}(a) are three times smaller than those from Ref.~\cite{Bzdak:2011yy} in Fig.~\ref{fig:EMFluctuation}(b) for central collisions. This discrepancy arises from differences in the nuclear thickness assumptions used in the calculations—Ref.~\cite{Deng:2012pc} accounted for a finite nuclear thickness, whereas Ref.~\cite{Bzdak:2011yy} assumed infinitely thin nuclei, incorporating Lorentz contraction at $\sqrt{s_{NN}}=$ 200 GeV.

\begin{widetext}
\begin{figure}[htbp]
\vspace*{-0.1in}
\includegraphics[scale=0.5]{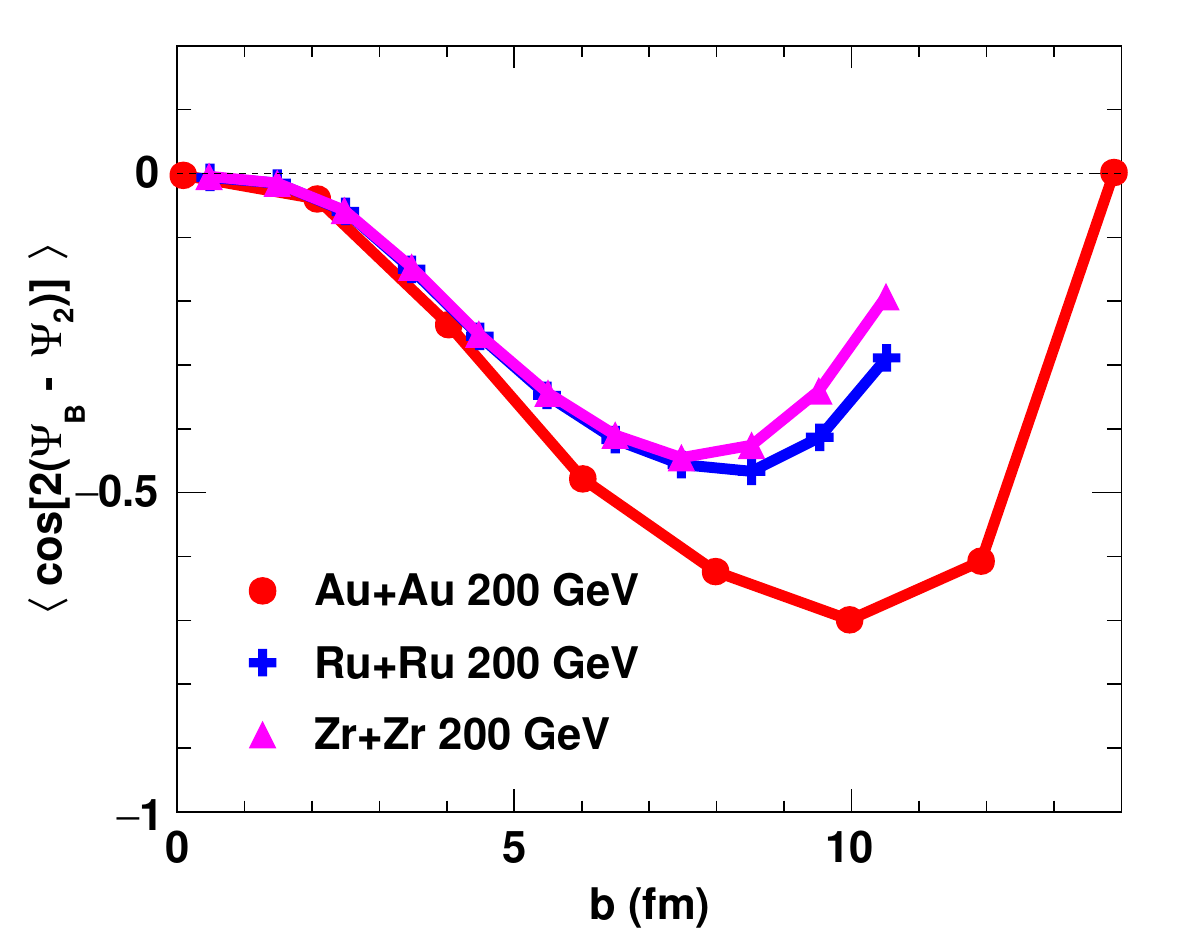}
\captionof{figure}{Correlations between magnetic field and second-harmonic participant plane in Au+Au, Ru+Ru, and Zr+Zr collisions at $\sqrt{s_{NN}}=$ 200 GeV (figure is from Ref.~\cite{Zhao:2019crj,Bloczynski:2012en}). The more negative $\langle \cos n(\Psi_B -\Psi_2)\rangle $ indicates a stronger correlation between the magnetic field and the participant plane $\Psi_2$.}
\label{fig:MagneticCorrelation}
\end{figure}
\end{widetext}

The event-by-event magnetic field plays a crucial role in electromagnetic effects within the QGP, as it remains nonzero even in the most central collisions due to fluctuations in the spatial positions of protons. For instance, the CME describes charge separation on an event-by-event basis and depends only on the magnetic field of the specific event. Experimentally, CME signals are analyzed along directions reconstructed from final-state particles, conventionally referred to as event planes. A nonzero magnetic field in the most central collisions could influence our interpretation of CME measurements as a function of centrality. Refs.~\cite{Zhao:2019crj, Bloczynski:2012en} investigated the correlations between the direction of the magnetic field and the event planes across different impact parameters. They found that the correlations between the magnetic field and the second-harmonic participant plane are significantly suppressed in both very central and very peripheral collisions, while peaking in mid-central collisions, as illustrated in Fig.~\ref{fig:MagneticCorrelation}.

\subsection{Time evolution of the magnetic field}
\begin{widetext}
\begin{figure}[htbp]
\vspace*{-0.1in}
\includegraphics[scale=0.25]{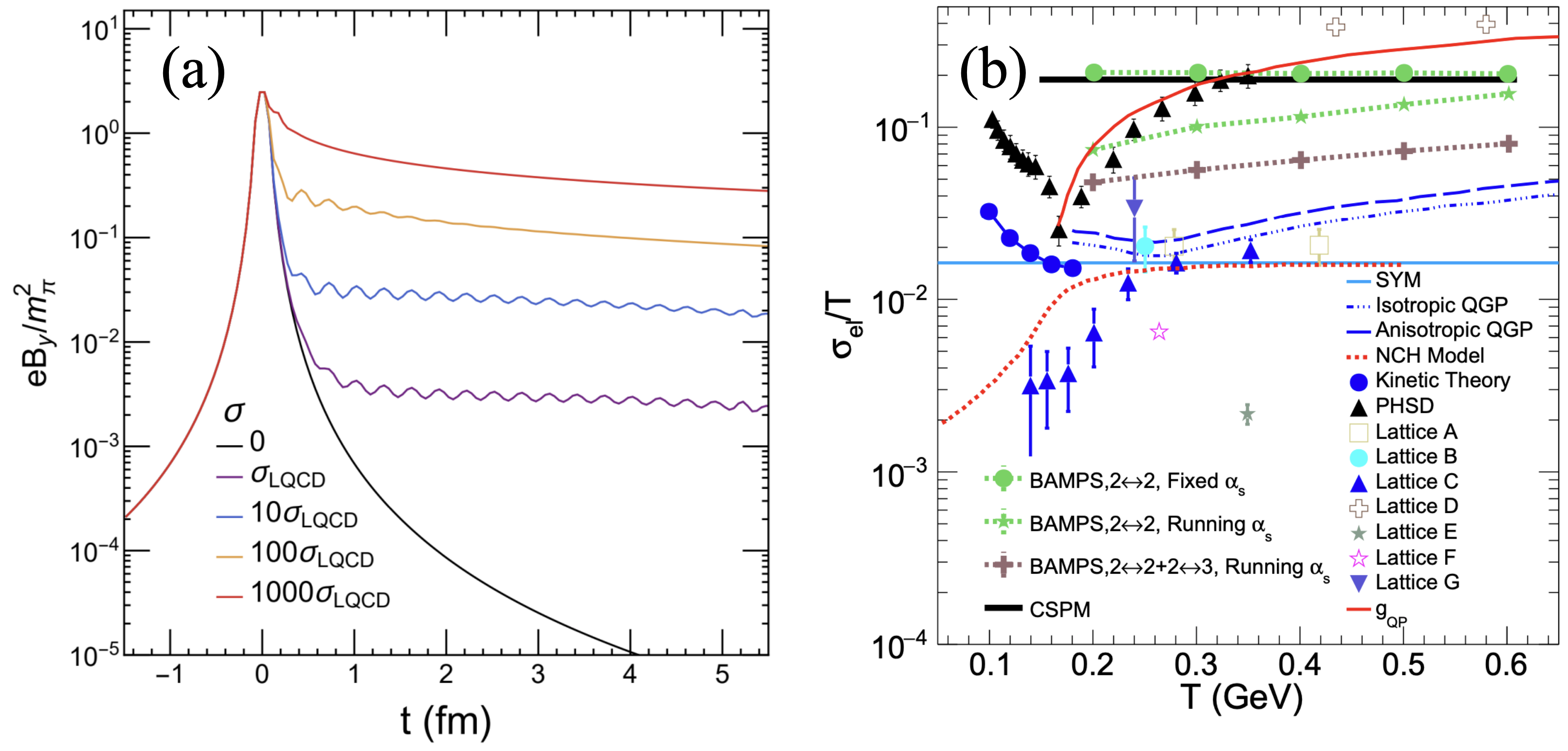}
\captionof{figure}{(a) The magnetic field evolution with different electrical conductivities in Au+Au collisions at $\sqrt{s_{NN}}=$ 200 GeV with impact parameter $b=$ 6 fm (figure is from Ref.~\cite{Huang:2022qdn}). (b) Estimates of the QGP's electrical conductivity as a function of temperature~\cite{Sahoo:2018dxn,Cassing:2013iz,Greif:2014oia,Finazzo:2013efa,Greif:2016skc,Caron-Huot:2006pee,Gupta:2003zh,Aarts:2007wj,Buividovich:2010tn,Ding:2010ga,Burnier:2012ts,Amato:2013naa,Thakur:2017hfc,Puglisi:2014pda} (figure is from Ref.~\cite{Sahoo:2018dxn}).}
\label{fig:MagneticEvolution}
\end{figure}
\end{widetext}

While the electromagnetic field can be well estimated using the Liénard-Wiechert potentials, its time evolution remains an open question due to the complex interplay with the QGP. A time-varying magnetic field induces an electric field through Faraday induction, which, in turn, generates an electric current that counteracts the decay of the external magnetic field in an electrically conductive medium—a phenomenon known as Lenz's law. Figure~\ref{fig:MagneticEvolution} (a) illustrates the evolution of the magnetic field in a static medium with different constant electrical conductivities~\cite{Huang:2022qdn}, demonstrating that medium response significantly prolongs the field's lifetime. However, theoretical estimates of the QGP's electrical conductivity have large uncertainties, as it depends on the medium's temperature. The conductivity-to-temperature ratio, $\sigma_{\rm el}/T$, extracted from lattice QCD~\cite{Buividovich:2010tn,Ding:2010ga,Burnier:2012ts,Amato:2013naa}, transport approaches~\cite{Cassing:2013iz,Greif:2014oia}, and effective models~\cite{Finazzo:2013efa,Sahoo:2018dxn}, spans two orders of magnitude, ranging from 0.001 to 0.4 for temperature from $0.1$ to $0.6$ GeV, as shown in Fig.~\ref{fig:MagneticEvolution} (b). In particular, substantial discrepancies exist between lattice QCD results and transport model predictions. These differences primarily arise because electrical conductivity reflects a non-equilibrium property of the QGP and is highly sensitive to how microscopic interactions are modeled in transport frameworks. For comparison, the QGP's conductivity estimated from Parton-Hadron-String Dynamics is approximately 500 times larger than that of copper and silver at room temperature~\cite{Cassing:2013iz}.

Moreover, the relaxation time of the medium response is also critical to the time evolution of the magnetic field. In Ref.~\cite{Li:2023tsf}, the authors numerically solved Maxwell’s equations in the QGP with a time-varying electric conductivity using the Finite-Difference-Time-Domain method. They found that the early-stage magnetic field is significantly suppressed when the conductivity becomes nonzero after the collision, compared to the case of a constant conductivity. This suppression arises because the medium takes time to develop its effect on delaying the decay of the magnetic field. However, at later times, the magnetic fields in both cases converge and remain larger than in the vacuum scenario, as illustrated in Fig.~\ref{fig:EarlyMagneticField} (a). 
The incomplete QGP electromagnetic response has also been studied using the Drude model with different relaxation times~\cite{Wang:2021oqq}. It was found that the relaxation time could be comparable to the lifetime of the external magnetic field, given typical values of conductivity and temperature in the QGP. As a result, the magnetic field at the early stage of the collision is suppressed by two orders of magnitude, as indicated by the solid lines in Fig.~\ref{fig:EarlyMagneticField} (b). 
Additionally, the pre-equilibrium properties of the QGP play a decisive role in the magnetic field evolution, as studied within a kinetic framework by solving the coupled Boltzmann and Maxwell equations~\cite{Yan:2021zjc}. The results suggest a rapid drop in the magnetic field strength during the pre-equilibrium stage of the QGP, as shown in Fig.~\ref{fig:EarlyMagneticField} (c). Furthermore, Ref.~\cite{Yan:2021zjc} found that while a residual magnetic field persists in the equilibrated QGP at RHIC energies, it becomes negligible at the LHC energies.

\begin{widetext}
\centering
\begin{figure*}[htbp]
\vspace*{-0.1in}
\includegraphics[scale=0.3]{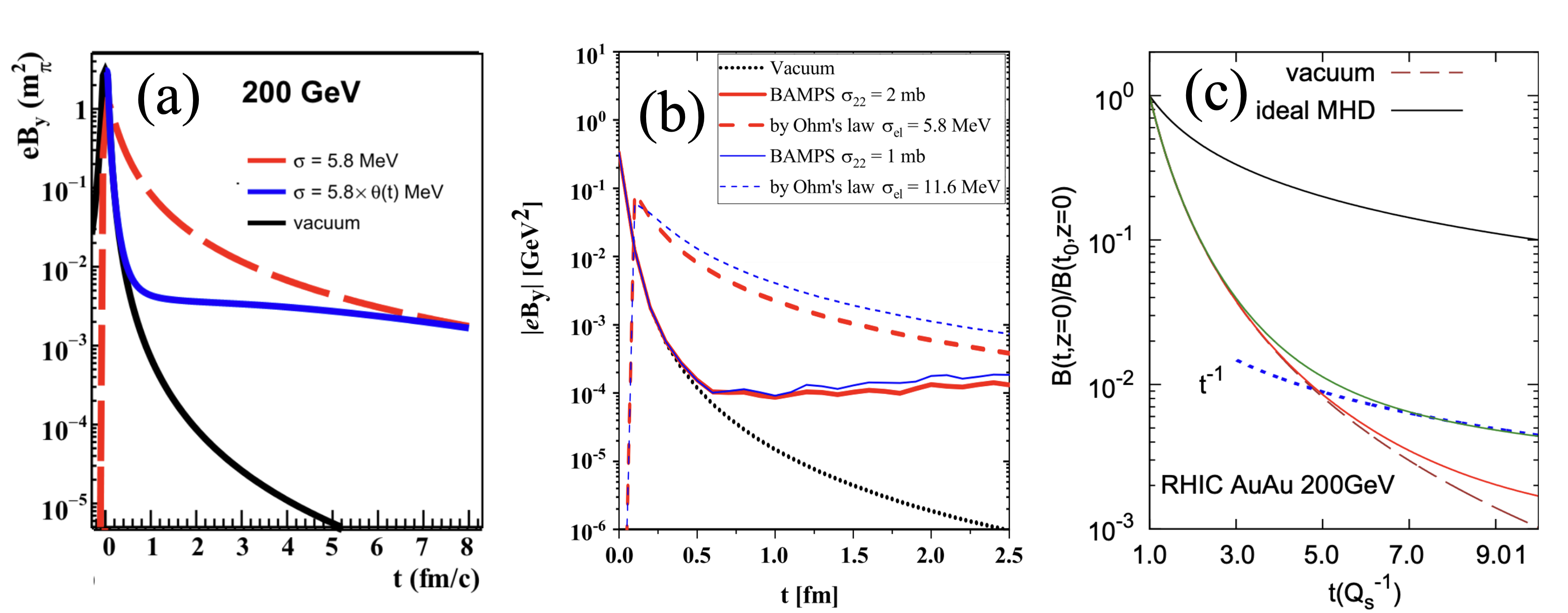}
\captionof{figure}{The magnetic field evolution with (a) time-dependent conductivity (blue solid line, figure is from Ref.~\cite{Li:2023tsf}), (b) incomplete QGP electromagnetic response (blue and red solid lines, figure is from Ref.~\cite{Wang:2021oqq}), and (c) in pre-equilibrium QGP (green solid line, figure is from Ref.~\cite{Yan:2021zjc}) in Au+Au collisions at $\sqrt{s_{NN}}=200$ GeV.}
\label{fig:EarlyMagneticField}
\end{figure*}
\end{widetext}

\begin{figure}[htbp]
\vspace*{-0.1in}
\includegraphics[scale=0.25]{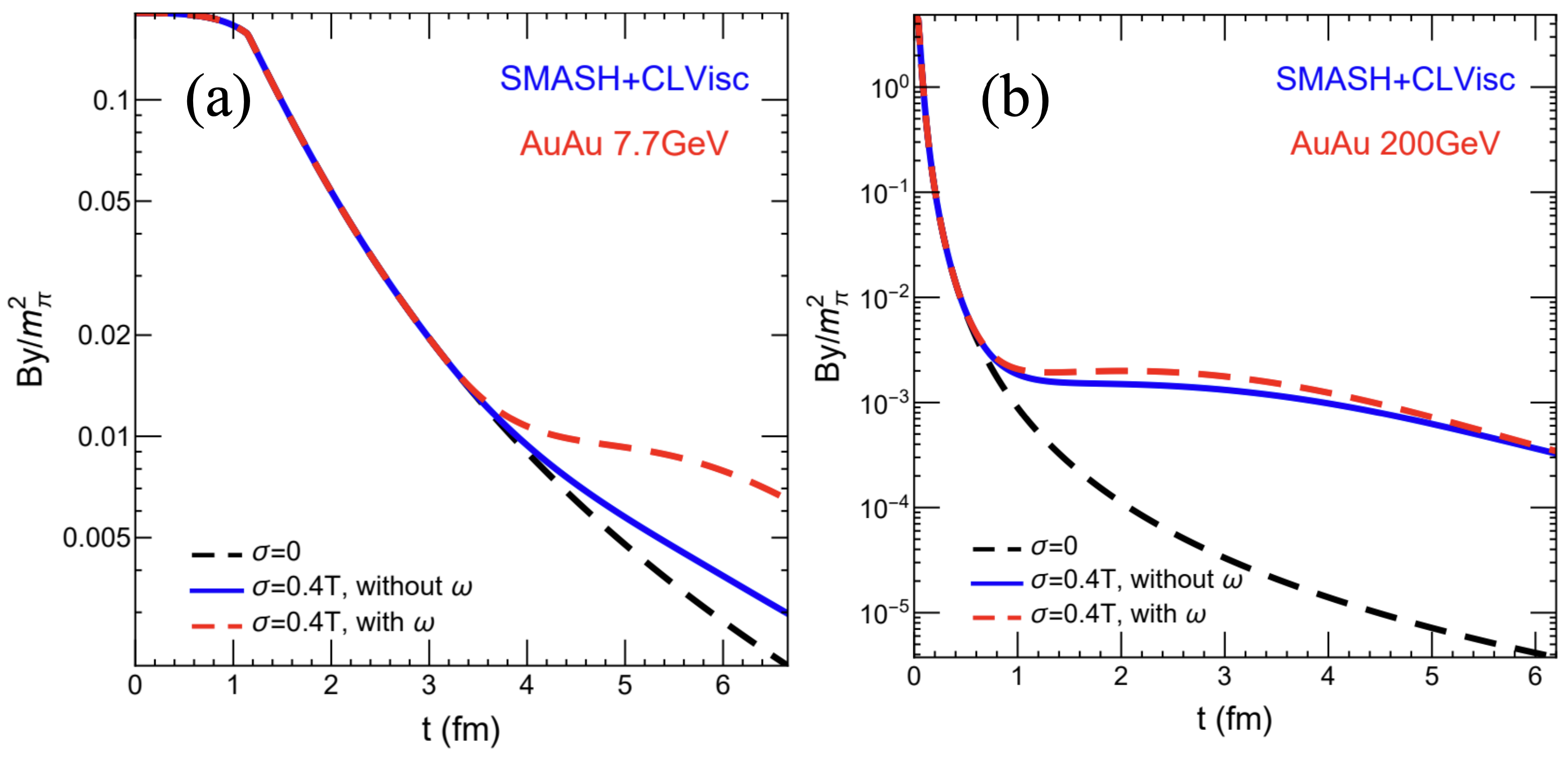}
\captionof{figure}{The evolution of the magnetic field with and without vorticity contribution in Au+Au collisions at $\sqrt{s_{NN}}=$ (a) 7.7 GeV and (b) 200 GeV, respectively (figure is from Ref.~\cite{Huang:2024aob}).}
\label{fig:VorticityEffect}
\end{figure}

In addition, the QGP produced in non-central heavy-ion collisions is the most vortical fluid observed in nature, with a vorticity of approximately $10^{21}$ $\rm s^{-1}$ at RHIC energies~\cite{STAR:2017ckg}. Recent studies suggest that this swirling QGP can induce an additional magnetic field aligned with the external field, effectively prolonging the lifetime of the magnetic field within the QGP~\cite{Huang:2024aob}. This effect is more pronounced at lower collision energies, as illustrated in Fig.~\ref{fig:VorticityEffect}, which compares Au+Au collisions at $\sqrt{s_{NN}}=$ 7.7 GeV and 200 GeV. 

\section{Experimental Results of The Electromagnetic Field Effects}\label{sec:exp}

\begin{figure}[htbp]
\vspace*{-0.1in}
\includegraphics[scale=0.4]{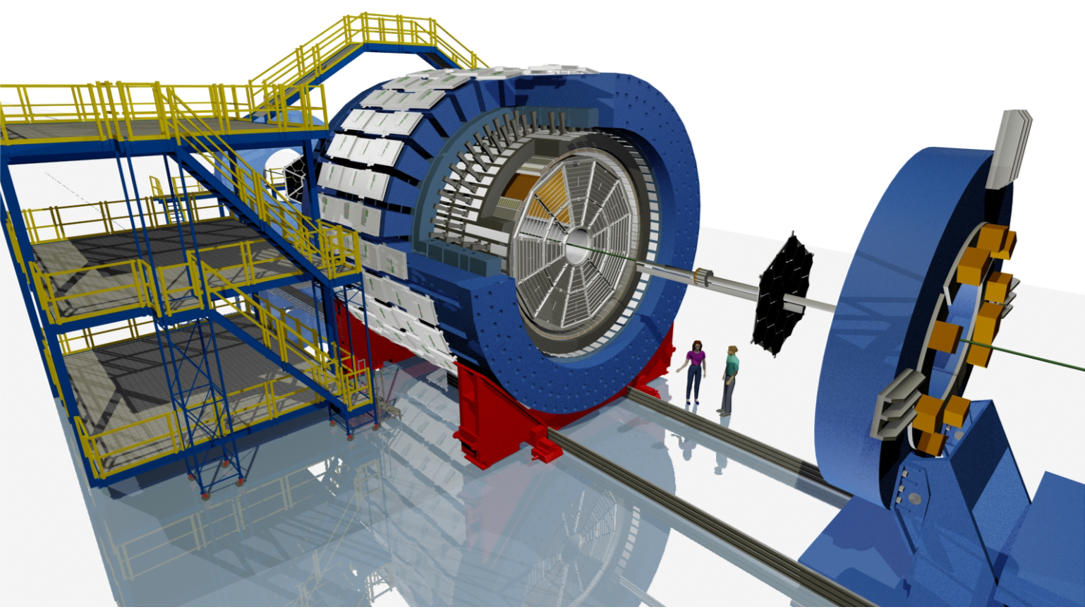}
\captionof{figure}{Perspective view of the STAR detector (figure is from the STAR collaboration).}
\label{fig:STARCartoon}
\end{figure}

Experimental data can help constrain theoretical estimates of the QGP's complex electromagnetic response by analyzing the electromagnetic field's imprints on final-state particles detected in the experiment. Detectors in heavy-ion collision experiments are typically complex apparatuses composed of multiple subsystems, designed to capture collision events with maximum clarity. For example, the Solenoidal Tracker at RHIC (STAR) has a diameter of approximately 15 meters to accommodate various subsystems for tracking, identifying, and measuring particles produced in collisions~\cite{ACKERMANN2003624}. Figure~\ref{fig:STARCartoon} sketches the STAR detector, with engineers on the ground for scale. Detecting a magnetic field with a lifetime on the order of $10^{-24}$s using a traditional magnetometer is extremely challenging. In heavy-ion collisions, the magnetic field's properties are inferred from the four-momenta of final-state particles. By employing observables sensitive to different stages of the magnetic field, its evolution can be experimentally investigated. However, some of these observables, such as those related to the chiral magnetic effect and heavy-flavor particles, require very large data samples due to their statistical demands. This section reviews measurements of effects induced by electromagnetic fields in heavy-ion collisions.

\subsection{Ultra-peripheral collisions}

\begin{figure}[htbp]
\vspace*{-0.1in}
\includegraphics[scale=0.3]{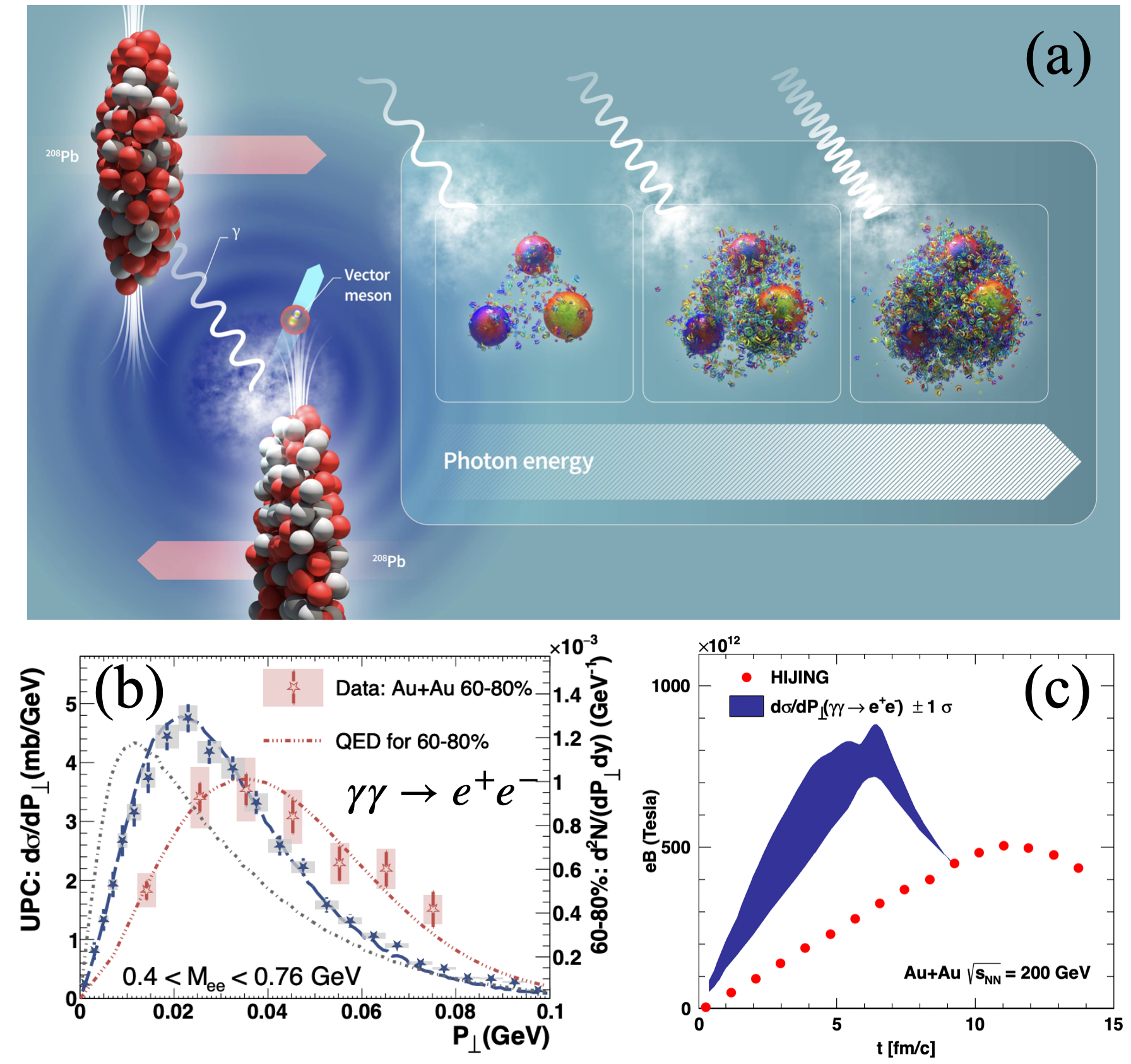}
\captionof{figure}{(a) Sketch of an ultra-peripheral collision (figure is from CERN). (b) Di-electron ${\bf P}_\perp$ distribution in UPC at RHIC (figure is from Ref.~\cite{STAR:2019wlg}). (c) The estimated magnetic field strength~\cite{Brandenburg:2021lnj} (blue band) and the HIJING model calculations~\cite{Deng:2012pc}.}
\label{fig:UPCB}
\end{figure}

In UPC, the impact parameter exceeds twice the nuclear radius~\cite{Bertulani:1987tz,Natale:2001zj}, as illustrated in Fig.~\ref{fig:UPCB} (a). Instead of engaging in strong interactions, the nuclei interact through their electromagnetic fields, which can be described as a high flux of energetic quasi-real photons~\cite{Baur:2001jj,Bertulani:2005ru,STAR_UPC,Ma_NST_2023,Shao_SCPMA,Ma_SCPMA_2024}. Elastic scattering between photons in UPC has been observed at the LHC~\cite{ATLAS:2017fur}, while di-electron production from inelastic photon-photon processes has been reported by RHIC experiments~\cite{STAR:2019wlg}, shown in Fig.~\ref{fig:UPCB}(b). These results confirm the generation of an electromagnetic field by fast-moving ions. The transverse momentum distribution of produced di-electron pairs, ${\bf P}_\perp$, in UPC, is sensitive to the spatial charge configuration of nuclei. Therefore, measuring di-electron ${\bf P}_\perp$ can help constrain the strength of the magnetic field generated by the charged protons within nuclei~\cite{Brandenburg:2021lnj}. Figure~\ref{fig:UPCB} (c) presents the magnetic field strength from fitting nuclear charge radius and skin depth extracted from experimental di-electron measurements~\cite{STAR:2019wlg}, the magnitudes are consistent with theoretical calculations using the HIJING model~\cite{Deng:2012pc}. 

\subsection{Charge-dependent directed flow as the witness of the electromagnetic field}
\begin{figure}[htbp]
\vspace*{-0.1in}
\includegraphics[scale=0.4]{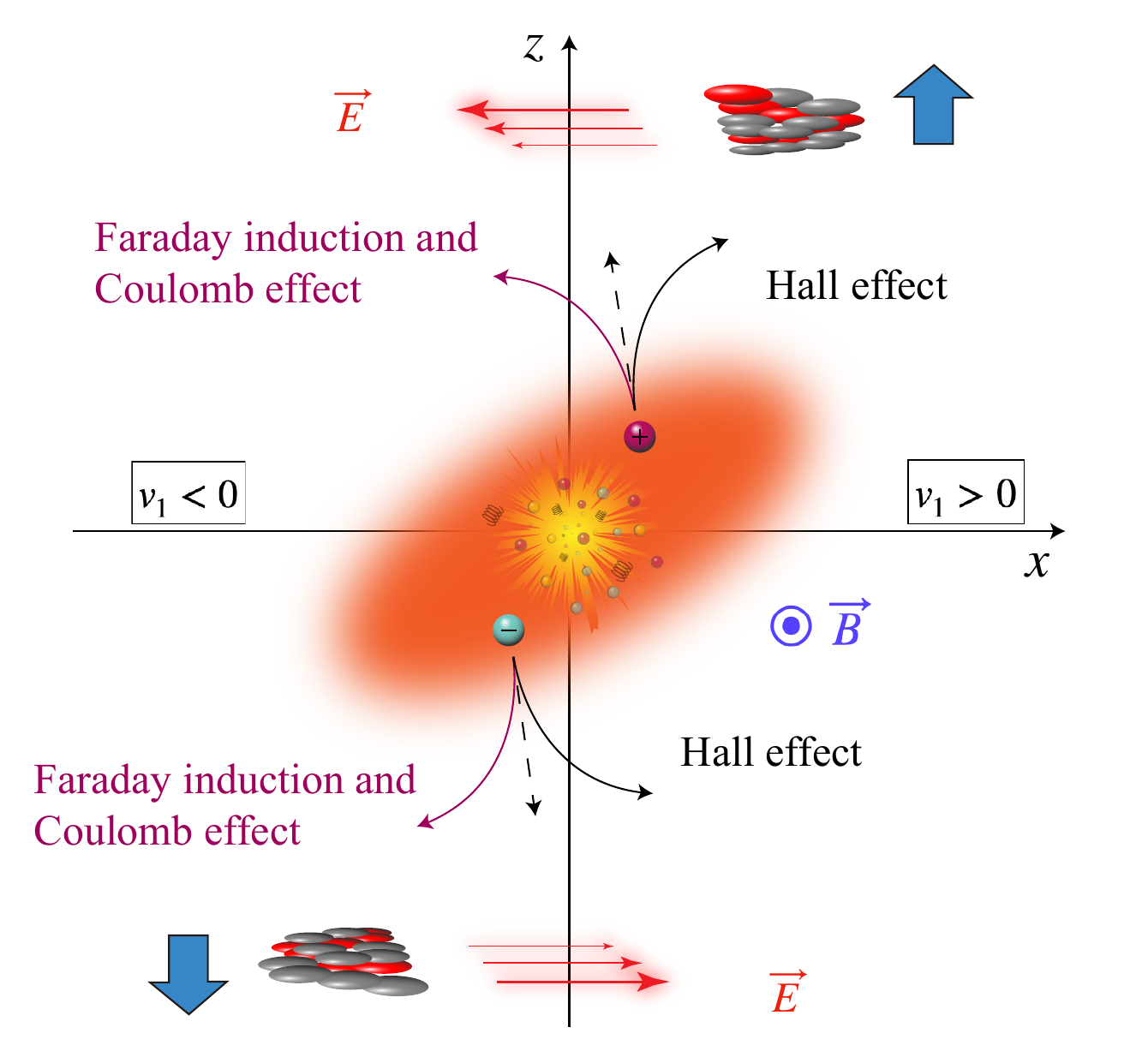}
\captionof{figure}{Hall effect, Faraday induction and Coulomb effect on the momenta of charged particles in a collision (figure is from Ref.~\cite{STAR:2023jdd}).}
\label{fig:EMforce}
\end{figure}

Many novel phenomena in heavy-ion collisions involve the magnetic field in the QGP, such as the CME and the chiral phase transition under a strong magnetic field. The charge-dependent collective motion of final-state particles can serve as a probe of the magnetic field in the medium, as charged particles in an electromagnetic field experience Lorentz and Coulomb forces, along with Faraday induction in the QGP, which collectively modify their momenta~\cite{Gursoy:2014aka,Gursoy:2018yai,Das:2016cwd,Nakamura:2022ssn,Sun:2023adv,Mohapatra:2011ku,Voronyuk:2014rna}. Figure~\ref{fig:EMforce} shows an overhead view of a heavy-ion collision. From this perspective, the system expansion influences the motion of oppositely charged particles similarly at leading order, while the electromagnetic fields introduce additional charge-dependent effects.
The azimuthal angle ($\phi$) distribution can be described by the following Fourier decomposition,
\begin{equation}
 \frac{dN}{d\Delta \phi} \propto 1+ \sum_{n=1}^{\infty} 2v_n\cos n \Delta\phi,
 \label{Eq:Fourier}
\end{equation}
where $\Delta \phi$ is the azimuthal angle of a particle with respect to the reaction plane (spanned by the $x$ and $z$ axis as shown in Fig.~\ref{fig:EMforce}), and $v_n$ is the $n{\rm th}$-harmonic flow coefficient. Directed flow $v_1$, arising from the initial tilt of the system, is an odd function of rapidity (y) and is sensitive to the rapidity-odd electromagnetic field generated by spectator protons~\cite{Gursoy:2014aka,Gursoy:2018yai,Das:2016cwd,Nakamura:2022ssn,Sun:2023adv}.  

The Hall effect from the Lorentz force, illustrated by the black lines in Fig.~\ref{fig:EMforce}, increases $v_1$ for a positive charge moving forward (y $>$ 0) and decreases it when moving backward (y $<$ 0), enhancing the gradient $dv_1/d$y. For a negative charge, the effect is reversed, reducing $dv_1/d$y. Consequently, the difference $\Delta dv_1/d$y between positive and negative charges is positive.
Additionally, the rapid decay of the magnetic field in the medium induces an electric field via Faraday induction. Combined with the electric field from spectator protons, this results in a negative $\Delta dv_1/d$y between positive and negative charges, as shown by the purplish-red lines. Theoretical calculations~\cite{Gursoy:2014aka,Gursoy:2018yai,Nakamura:2022ssn} suggest that the combined effects of Faraday induction and Coulomb interactions are expected to dominate over the Hall effect for light quarks that are assumed to be in thermal equilibrium. On the other hand, for charm quarks, which form earlier and remain out of equilibrium, the effects are reversed~\cite{Das:2016cwd}. Ultimately, these interactions will lead to a measurable difference in $v_1$ between positively and negatively charged hadrons via quark coalescence.

\begin{figure}[htbp]
\vspace*{-0.1in}
\includegraphics[scale=0.5]{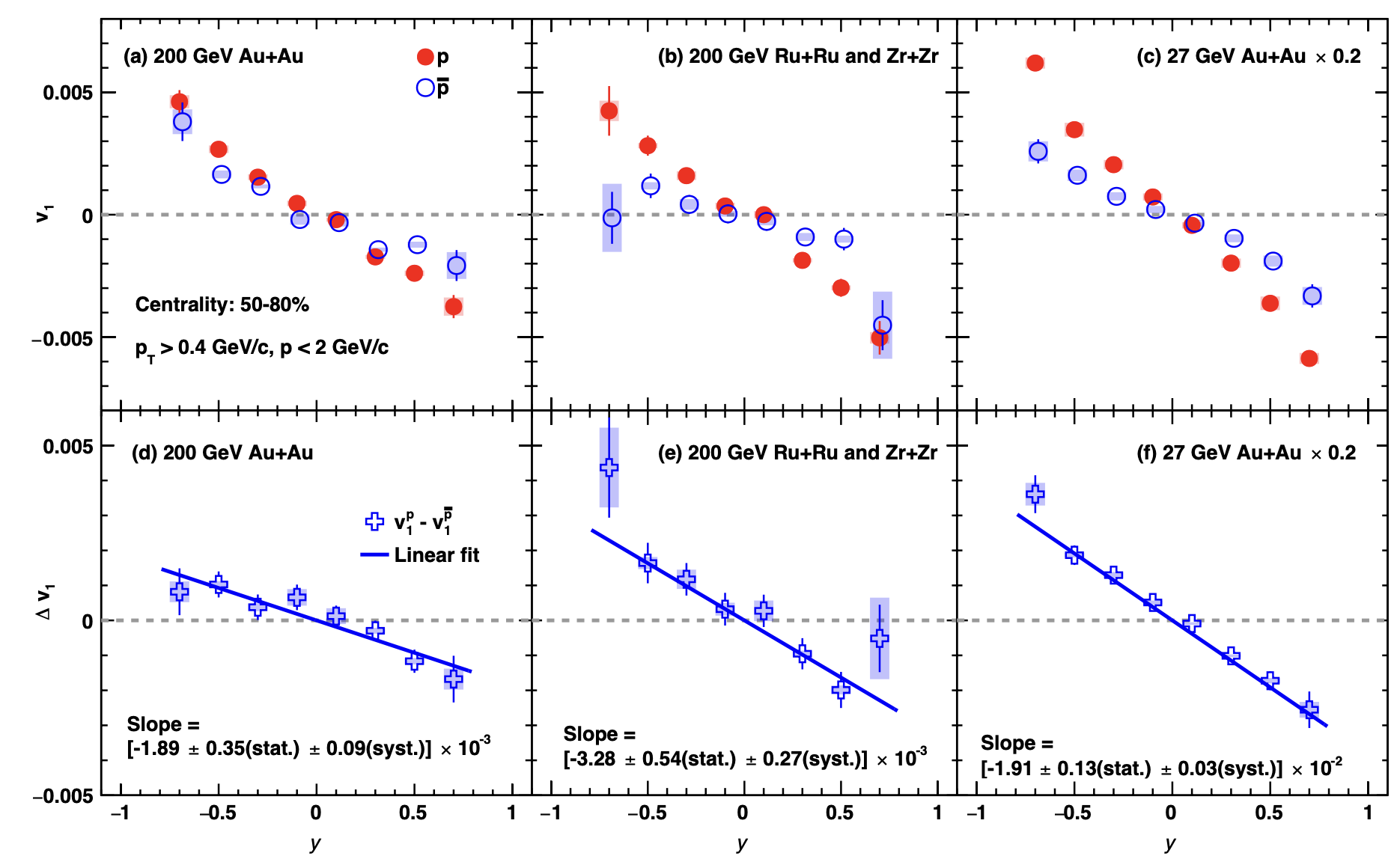}
\captionof{figure}{(Upper) Proton and anti-proton directed flow in Au+Au, Ru+Ru and Zr+Zr collisions at $\sqrt{s_{NN}}=$ 200 GeV, and Au+Au collisions at $\sqrt{s_{NN}}=$ 27 GeV in the 50--80\% centrality range (figure is from ~\cite{STAR:2023jdd}). (Lower) The difference $\Delta dv_1/d$y between protons and anti-protons.}
\label{fig:protonv1}
\end{figure}

\begin{figure}[htbp]
\vspace*{-0.1in}
\includegraphics[scale=0.4]{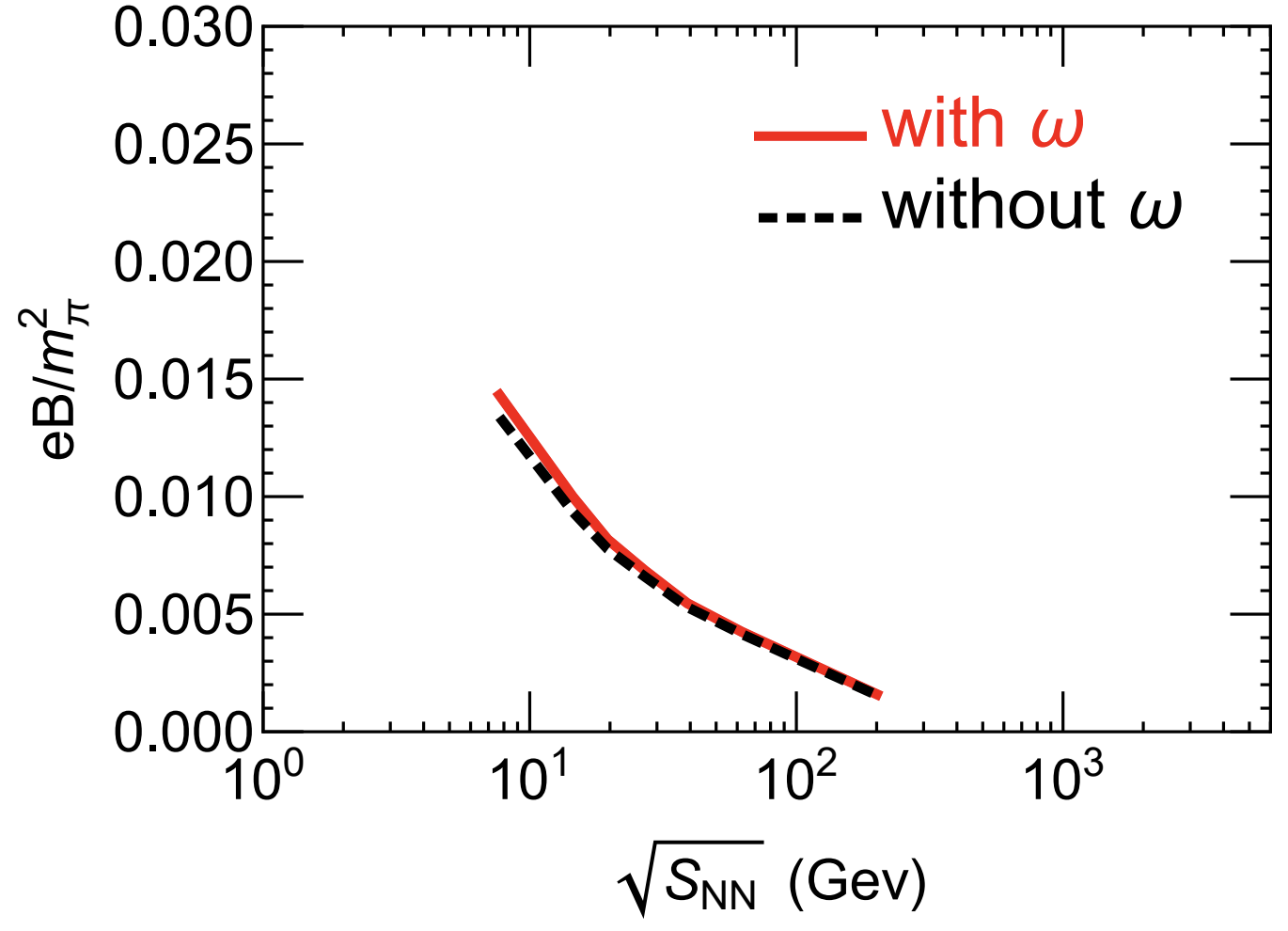}
\captionof{figure}{The magnetic field strength on the freeze-out hypersurface at different collision energies (figure is from Ref.~\cite{Huang:2024aob}).}
\label{fig:lateB}
\end{figure}

Recent $v_1$ measurements for protons and anti-protons exhibit a negative $\Delta dv_1/d$y, consistent with expectations from Faraday induction and the Coulomb effect, in the 50–-80\% centrality range for Au+Au, Ru+Ru, and Zr+Zr collisions at $\sqrt{s_{NN}}=$ 200 GeV, as well as for Au+Au collisions at $\sqrt{s_{NN}}=$ 27 GeV, as shown in Fig.~\ref{fig:protonv1}. For the first time in experiments, signals consistent with the electromagnetic field effect on the QGP have been observed in symmetric collisions, with significances exceeding 5$\sigma$. It is proposed that the $\Delta dv_1/d$y of light quarks, such as $u$ and $d$, are sensitive to the magnetic field at freeze-out~\cite{Gursoy:2014aka,Gursoy:2018yai}. The more negative $\Delta dv_1/d$y at $\sqrt{s_{NN}} = 27$ GeV compared to that at $\sqrt{s_{NN}}=$ 200 GeV suggests a stronger late-stage magnetic field at lower energies, likely due to the longer passage time of incident nuclei and the shorter lifetime of the QGP.
Ref.~\cite{Huang:2024aob} examines the magnetic field strength on the QGP freeze-out hypersurface at several collision energies available at RHIC, and finds that the strength of the late-stage magnetic field is higher at lower energies regardless of the vorticity effect, as shown in Fig.~\ref{fig:lateB}. This result can explain the more negative $\Delta dv_1/d$y measured at $\sqrt{s_{NN}} = 27$ GeV.   

\begin{figure}[htbp]
\vspace*{-0.1in}
\includegraphics[scale=0.5]{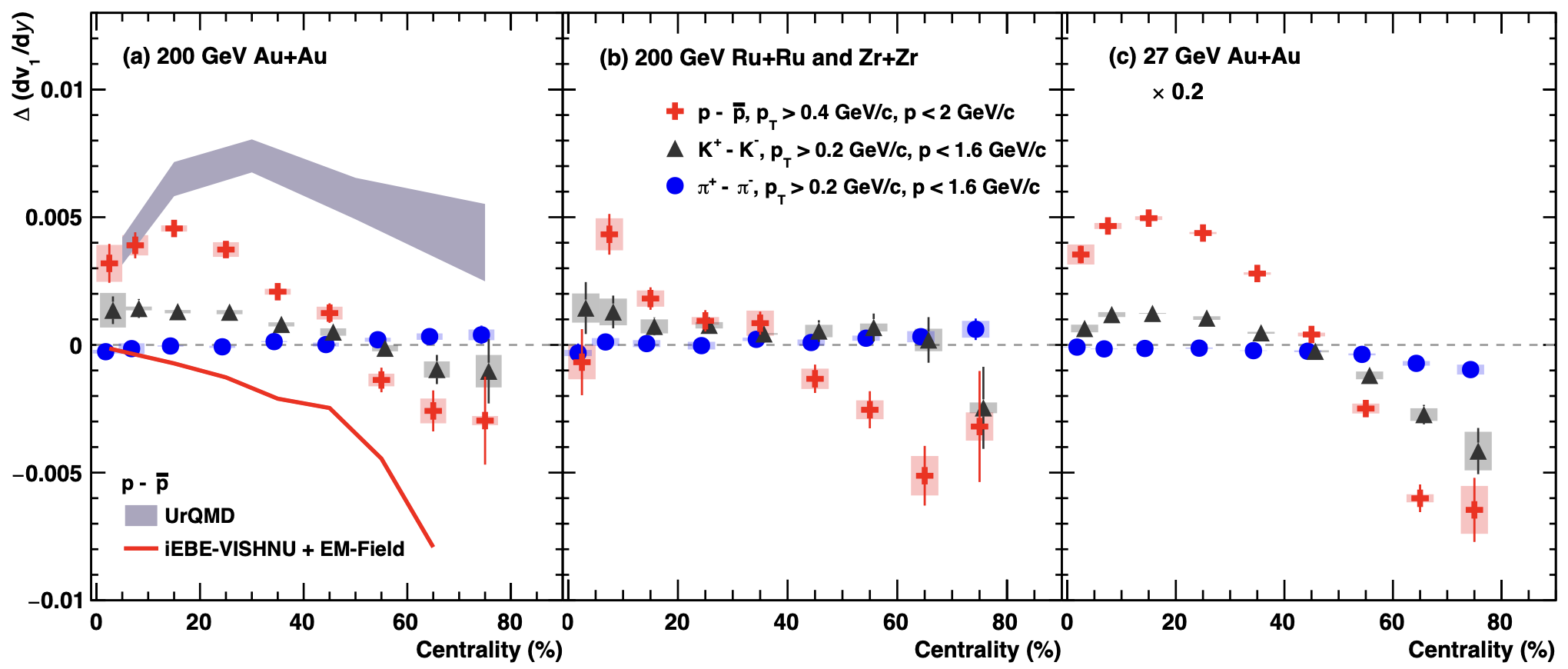}
\captionof{figure}{$\Delta dv_1/d$y between positively and negatively charged pions, kaons, and protons as a function of centrality in different systems (figure is from Ref.~\cite{STAR:2023jdd}).}
\label{fig:v1centrality}
\end{figure}

Besides electromagnetic effects, quarks transported from the incident nucleons can also influence the $\Delta dv_1/d$y between protons and anti-protons, as transported $u$ and $d$ quarks may contribute to proton formation through quark recombination. Different model studies~\cite{Guo:2012qi,Nayak:2019vtn,Bozek:2022svy} suggest that transported quarks give a positive contribution to proton $\Delta dv_1/d$y, hindering the observation of the electromagnetic effect on $\Delta dv_1/d$y in central collisions, where the magnetic field is weak. Figure~\ref{fig:v1centrality} shows the centrality dependence of $\Delta dv_1/d$y for pions, kaons, and protons. The $\Delta dv_1/d$y values for kaons and protons transition from positive in central collisions to negative in peripheral ones, reflecting the increasing dominance of magnetic field effects. A hydrodynamic model with an inhomogeneous baryon density as the initial condition reproduces the proton $\Delta dv_1/d$y but fails to simultaneously describe the pion and kaon data~\cite{Parida:2023ldu}.  

\begin{figure}[htbp]
\vspace*{-0.1in}
\includegraphics[scale=0.35]{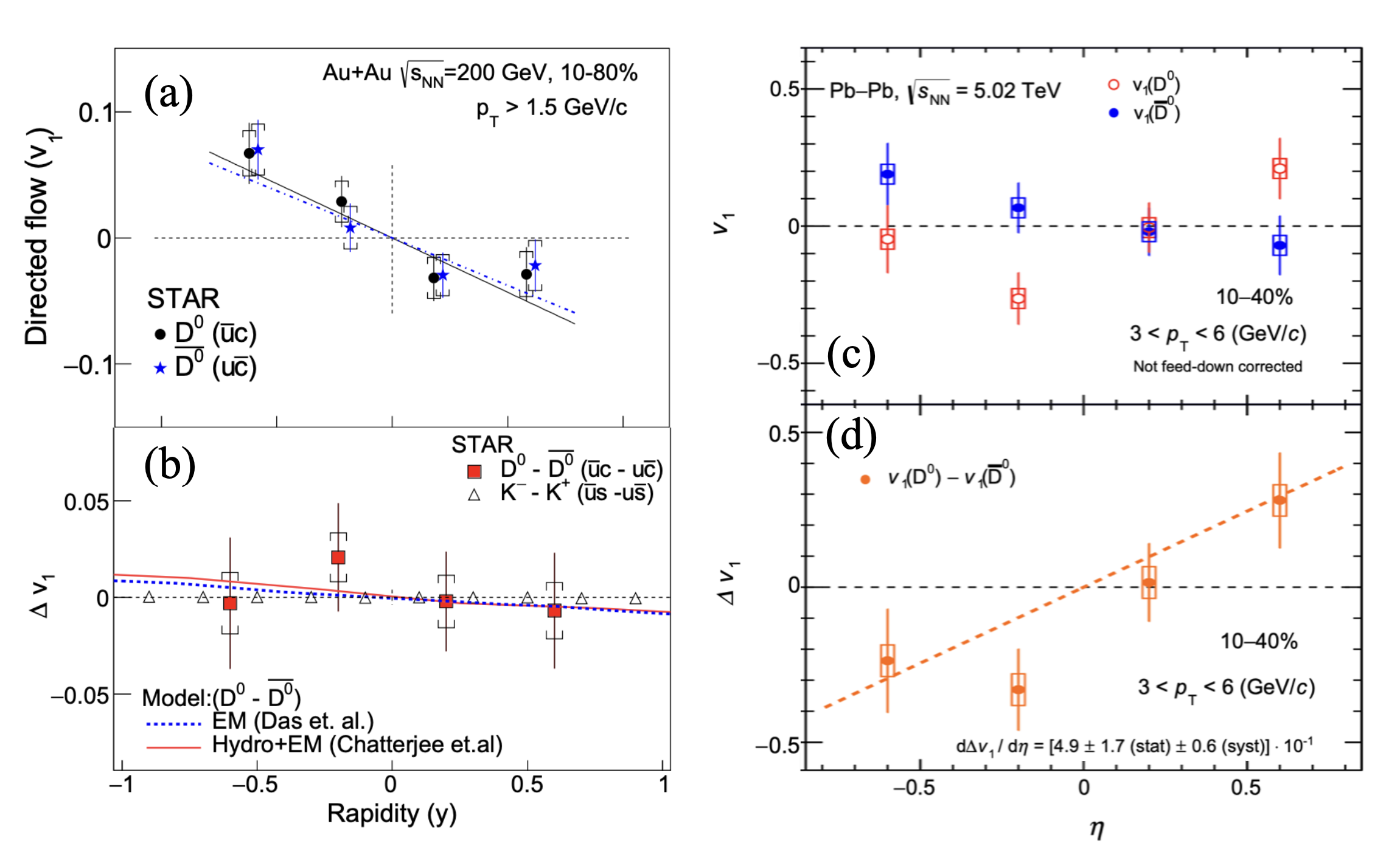}
\captionof{figure}{(a) $v_1$ for $D^0$ and $\bar{D^0}$, as well as (b) the difference vs rapidity in Au+Au collisions at $\sqrt{s_{NN}}=$ 200 GeV~\cite{STAR:2019clv}. (c) $v_1$ for $D^0$ and $\bar{D^0}$, as well as (d) the difference vs pseudo-rapidity in Pb+Pb collisions at $\sqrt{s_{NN}}=$ 5.02 TeV (figure is from Ref.~\cite{ALICE:2019sgg}).}
\label{fig:Dmeson}
\end{figure}

$D^0$ mesons are expected to be sensitive to the early-stage electromagnetic field, as charm quarks are heavy and produced early in the collisions~\cite{Das:2016cwd}. Since the magnetic field is strongest in the early stage, theoretical calculations suggest that the directed flow splitting between $D^0$ and $\bar{D^0}$ is primarily driven by the Hall effect. Figure~\ref{fig:Dmeson} shows the experimental results. Due to limited statistics, the measured $\Delta dv_1/d$y between $D^0$ and $\bar{D^0}$ in Au+Au collisions at $\sqrt{s_{NN}} = 200$ GeV is consistent with both zero and the calculation with the electromagnetic field effect~\cite{STAR:2019clv}. The results from Pb+Pb collisions at $\sqrt{s_{NN}} = 5.02$ TeV show a positive $\Delta dv_1/d$y, with a significance level below 3$\sigma$.

\begin{figure}[htbp]
\vspace*{-0.1in}
\includegraphics[scale=0.32]{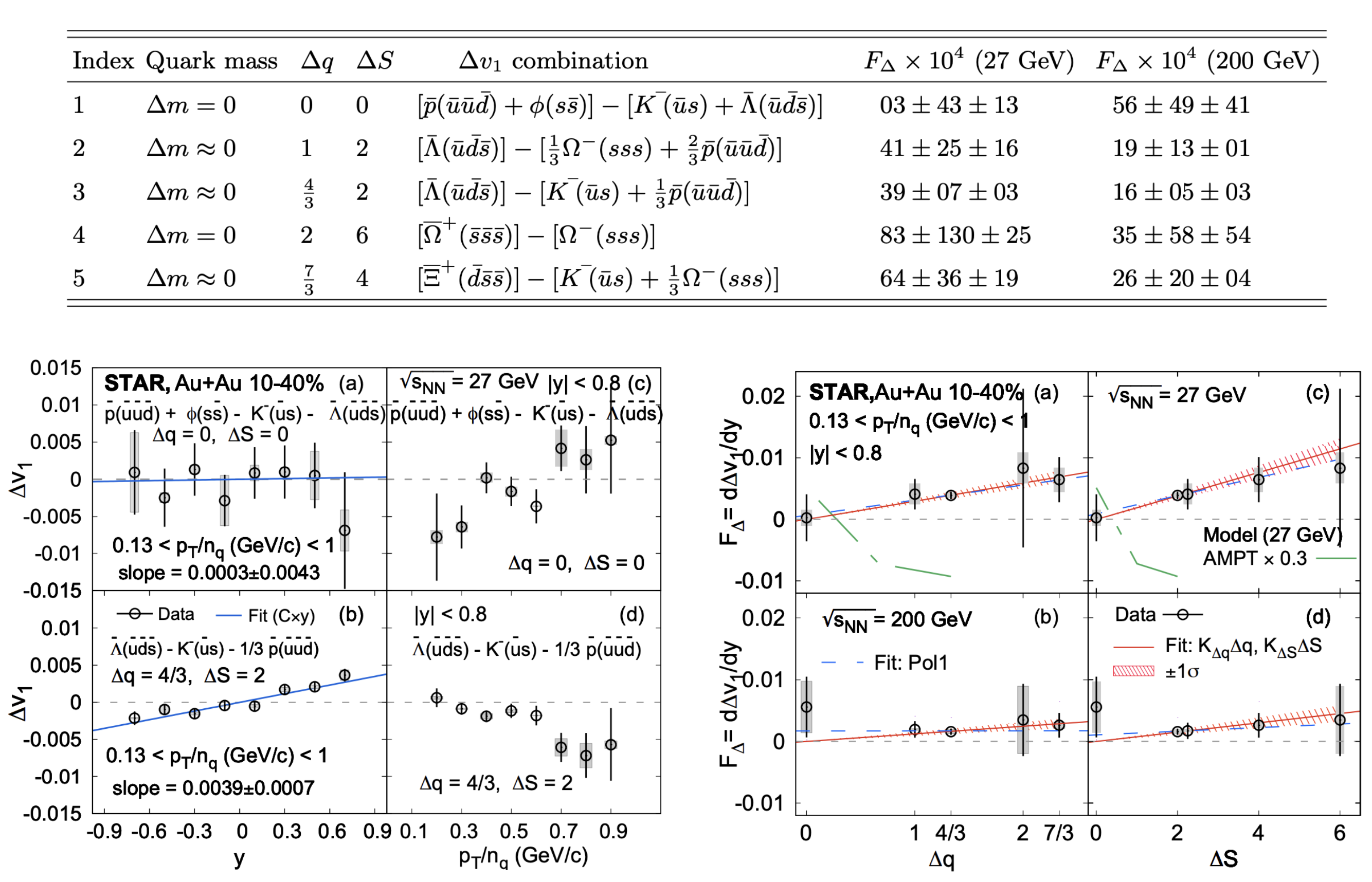}
\captionof{figure}{$\Delta dv_1/d$y between different hadron combinations containing no transported quarks in 10--40\% Au+Au collisions at $\sqrt{s_{NN}}=$ 27 GeV (figure is from Ref.~\cite{STAR:2023wjl}).}
\label{fig:v1Combination}
\end{figure}

To avoid the contribution from transported quarks, Ref.~\cite{STAR:2023wjl} examines combinations of hadrons that do not contain transported quarks and assumes the quark coalescence sum rule. Figure~\ref{fig:v1Combination} shows that in the 10--40\% centrality region, the measured $\Delta dv_1/d$y between different hadron combinations with the same quark masses increases with both the charge difference and the difference in strangeness number. From the perspective of the electromagnetic field effect, the results suggest that the Hall effect is dominant in more central collisions, or that it is the dominant effect for strange quarks at low energies. 

\begin{figure}[htbp]
\vspace*{-0.1in}
\includegraphics[scale=0.3]{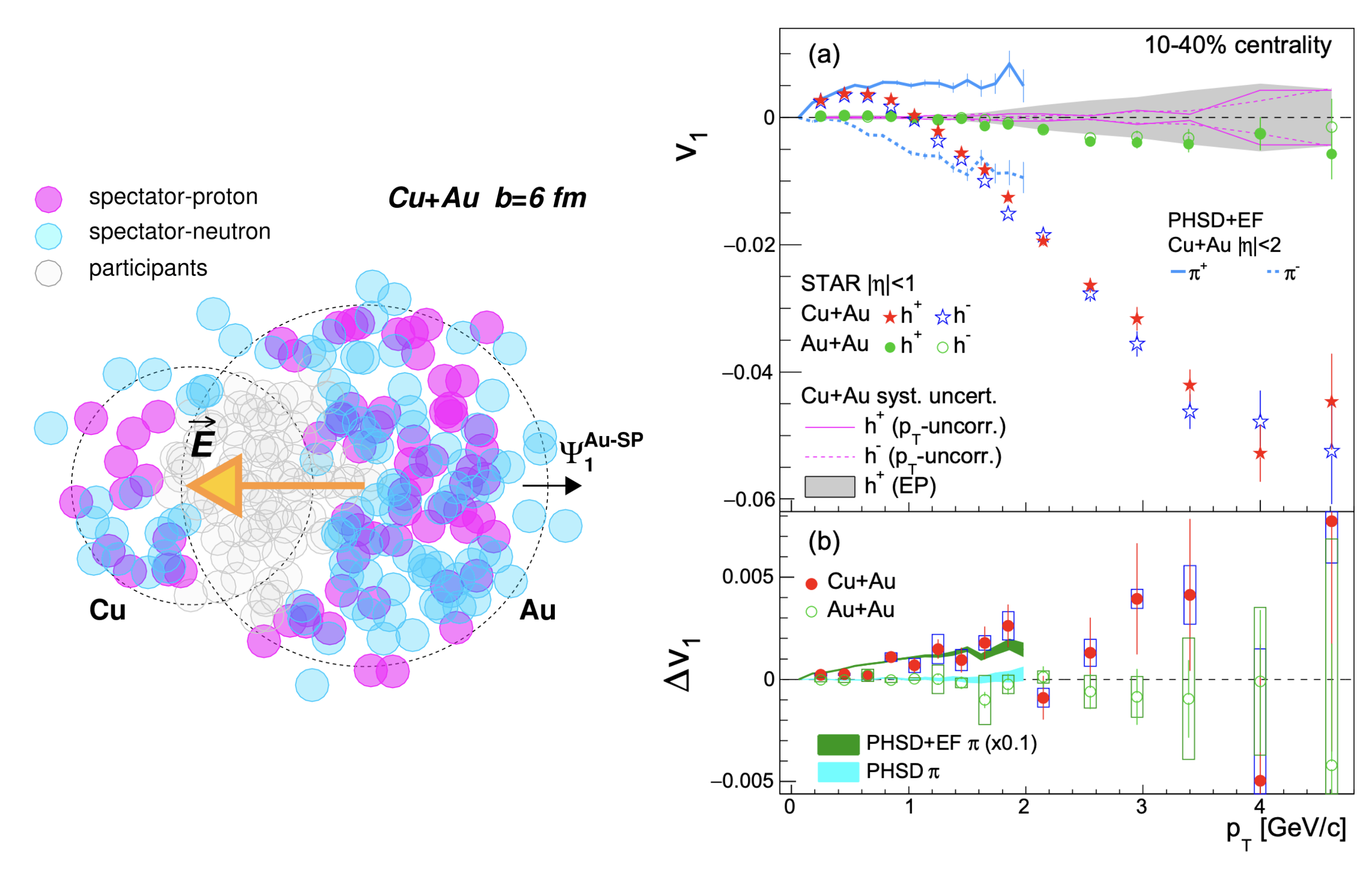}
\captionof{figure}{$v_1$ for positive and negative charges (a)
and the difference (b) in Cu+Au collisions at $\sqrt{s_{NN}}=$ 200 GeV (figure is from Ref.~\cite{STAR:2016cio}).}
\label{fig:v1CuAu}
\end{figure}

In an asymmetric collision, where one of the nuclei has more protons, a net Coulomb field is directed from the larger nucleus to the smaller one~\cite{Hirono:2012rt,Deng:2014uja,Voronyuk:2014rna}. This Coulomb field is an even function of rapidity and could generate a rapidity-even $\Delta dv_1/d$y between positively and negatively charged quarks~\cite{Voronyuk:2014rna}. Since the net electric field originates from spectators, it exists only in the very early stages of the collisions. The charge-dependent directed flow in asymmetric collisions can reveal the electromagnetic properties of the system before equilibrium, such as the production time of charged quarks during the collisions. Figure~\ref{fig:v1CuAu} shows the $\Delta dv_1/d$y between positively and negatively charged hadrons in Cu+Au collisions at $\sqrt{s_{NN}}=$ 200 GeV~\cite{STAR:2016cio}. The results are qualitatively consistent with the Coulomb effect, but the magnitudes are much smaller than the predictions from the parton-hadron-string dynamics model, suggesting that most of the quarks are produced at a later stage.

\subsection{Global polarization of $\Lambda$ ($\bar{\Lambda}$) under magnetic fields}
\begin{figure}[htbp]
\vspace*{-0.1in}
\includegraphics[scale=0.3]{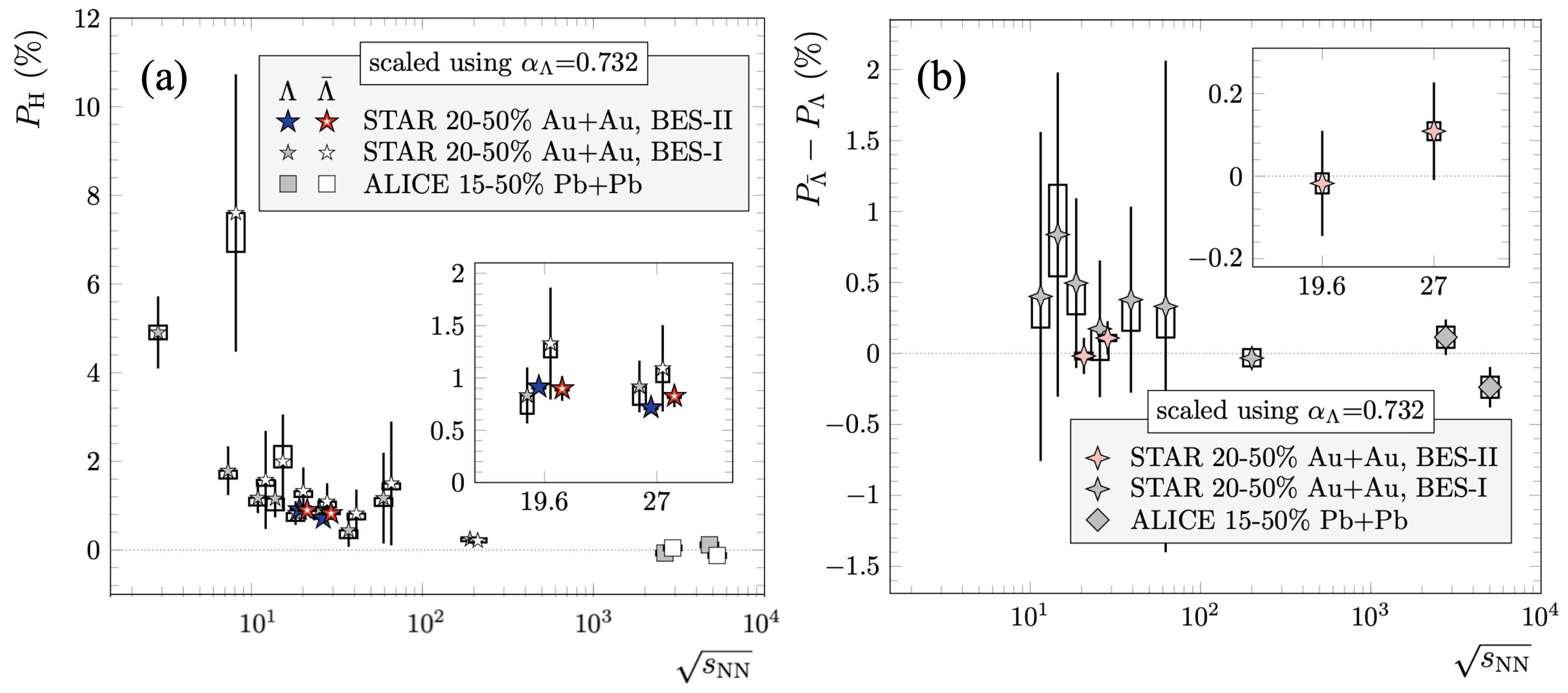}
\captionof{figure}{Global polarizations of $\Lambda$ and $\bar{\Lambda}$ (a) and the difference (b) in heavy-ion collisions~\cite{STAR:2023nvo,STAR:2017ckg,STAR:2018gyt,STAR:2021beb,ALICE:2019onw} (figure is from Ref.~\cite{STAR:2018gyt}).}
\label{fig:Lambda}
\end{figure}

Spin is an intrinsic property of particles, associated with the magnetic moment that interacts with external magnetic fields. Particles with opposite magnetic moments polarize in opposite directions, aligning with the external magnetic field. In heavy-ion collisions, besides the magnetic field, the substantial vorticity of the QGP can also induce global polarization through spin-orbital coupling in strong interactions~\cite{Liang:2004ph,Sun:2025oib,Chen:2023mel}. On average, the direction of the magnetic field aligns with the global orbital angular momentum, resulting in quark polarization in approximately the same direction. However, while the magnetic field distinguishes particles and antiparticles, spin-orbital coupling does not exhibit such differentiation. The global polarization or spin alignment effect has been observed for $\Lambda$, $\bar{\Lambda}$, and $\phi$ in Au+Au collisions at RHIC~\cite{STAR:2023nvo,STAR:2022fan,STAR:2017ckg,STAR:2018gyt,STAR:2021beb,STAR:2008lcm,Chen:2024afy,ChenJH-2023}, as well as in Pb+Pb collisions at the LHC~\cite{ALICE:2019onw,ALICE:2019aid,ALICE:2022dyy}. The measured polarizations for both hyperons and anti-hyperons are on the order of a few percent, with their difference, $\Delta P_{\Lambda} = P_{\Lambda} - P_{\bar{\Lambda}}$, remaining within 1\%, as shown in Fig.~\ref{fig:Lambda}. The magnetic field strength can be estimated via~\cite{Peng:2022cya,Muller:2018ibh,Xu:2022hql,Huang:2024aob}
\begin{equation}
    \Delta P_{\Lambda} = 2\mu_{\Lambda}\frac{B}{T_{\Lambda}},
\end{equation}
where the magnet moment of $\Lambda$ is $\mu_{\Lambda}=1.93\times 10^{-14}$ MeV/T, and its freeze-out temperature is assumed to be $T_{\Lambda}=$ 150 MeV. Based on the most accurate measurements from Au+Au collisions at $\sqrt{s_{NN}}=$ 19.6 GeV, 27 GeV~\cite{STAR:2023nvo}, and 200 GeV~\cite{STAR:2018gyt}, the upper limits of the late-stage magnetic field at a 95\% confidence level are $9.4 \times 10^{16}$ Gauss, $1.4 \times 10^{17}$ Gauss, and $6.0 \times 10^{16}$ Gauss for $\sqrt{s_{NN}}=$ 19.6 GeV, 27 GeV, and 200 GeV, respectively. Comparing with Fig.~\ref{fig:MagneticEvolution}, which illustrates the magnetic field evolution in Au+Au collisions at $\sqrt{s_{NN}}=$ 200 GeV, we can infer that the electrical conductivity of the QGP, as predicted by lattice QCD calculations, is of the right magnitude. Moreover, the beam energy dependence of $\Delta P_\Lambda$ provides a means to probe the lifetime of the magnetic field across different collision energies. Early results from the RHIC Beam Energy Scan (BES) Phase I suggest an inverse relationship between the magnetic field lifetime and the beam energy~\cite{Guo:2019joy}. With increased statistics from the RHIC BES-II program, the uncertainties in the energy dependence of $\Delta P_\Lambda$ can be further reduced. 

\section{Challenges and opportunities}\label{sec:cha}
Maxwell's equations provide the most successful framework for describing electromagnetic fields. Based on these equations, a very strong magnetic field would arise in relativistic heavy-ion collisions. The peak value of this magnetic field is determined by the spatial distribution and velocity of the electric charges involved. For a given collision energy and impact parameter, the magnetic field strength in the vacuum can be reliably estimated theoretically. Although the initial charge distribution depends on nuclear structure and carries some uncertainty, the peak magnetic field values in heavy-ion collisions are generally consistent across different assumptions about nuclear charge distributions~\cite{Voronyuk:2011jd,Skokov:2009qp,Kharzeev:2007jp,Zhao:2019crj}. However, the evolution of the magnetic field, particularly its interaction with the QGP, remains theoretically undetermined. In the early stages of the collision, different models predict significantly different magnetic field strengths, while their evolution curves tend to converge in the later stages, as illustrated in Fig.~\ref{fig:EarlyMagneticField}. 

Experimental measurements of the magnetic field effects can provide insights into its evolution. For instance, the directed flow of $D^0$ mesons can probe the early-stage magnetic field~\cite{Das:2016cwd}. However, early experimental results have large uncertainties, as shown in Fig.~\ref{fig:Dmeson}. The ALICE collaboration has collected several orders of magnitude more data in the Run 3 project compared with the previous $D^0$ meson $v_1$ publication, and results from this new dataset are highly anticipated. Meanwhile, the directed flow of light quarks is sensitive to the late-stage magnetic field~\cite{Gursoy:2014aka,Gursoy:2018yai,Nakamura:2022ssn}. The charge-splitting effect in light-hadron directed flow has shown qualitative agreement with theoretical calculations. However, quantitative comparisons with theoretical models require full simulations incorporating both transported-quark effects and magnetic-field effects, necessitating further theoretical work. Experimentally, measuring the charge-dependent directed flow of light hadrons at different collision energies can help explore the beam energy dependence of the magnetic field.

For the global polarization splitting between $\Lambda$ and $\bar{\Lambda}$, current results are consistent with both the magnetic field evolution based on lattice QCD conductivity and the scenario with no magnetic field. In Run 2025, the STAR collaboration will collect ten times more events from Au+Au collisions at $\sqrt{s_{NN}} = 200$ GeV. Results from this new dataset will provide stronger constraints on the late-stage magnetic field. 

Moreover, new observables have recently been proposed to study the magnetic field in experiments, such as baryon electric charge correlation~\cite{Ding:2023bft,Huang:NST} and elliptic flow of direct photons~\cite{Sun:2023rhh}. Experimental measurements of these effects are eagerly anticipated.

\section{summary}\label{sec:sum}
This review article explores the intense electromagnetic fields generated in heavy-ion collisions at relativistic energies, where accelerated nuclei create high-temperature environments leading to quark deconfinement. These fields, with strengths reaching $10^{18}$ Gauss at RHIC and $10^{19}$ Gauss at the LHC, can significantly impact the produced particles and serve as a tool to investigate the electromagnetic properties of the QGP. We have presented a comprehensive overview of the magnetic field generated in heavy-ion collisions, covering its generation, evolution, and experimental status. While the peak value of the magnetic field is relatively well constrained theoretically, its evolution within the QGP remains an open question. Experimental measurements of magnetic field effects at various stages are essential for mapping out this evolution. For instance, the directed flow of $D^0$ mesons is believed to be sensitive to the early-stage magnetic field. Although early measurements at the LHC suffer from large uncertainties, more precise $D^0$ meson $v_1$ data from the Run 3 program are expected to provide deeper insights into the early magnetic field dynamics. The article consolidates existing knowledge and serves as a comprehensive resource to compile most aspects of the magnetic field in this context. Additionally, we suggest future directions for further research in this area of nuclear and high-energy physics.

\section*{Acknowledgement}
This work is supported in part by the National Key Research and Development Program of China under Contract No. 2022YFA1604900, by the National Natural Science Foundation of China (NSFC) under Contract Nos. 12147101, 12205050, 12025501, and 12225502, the Guangdong Major Project of Basic and Applied Basic Research No. 2020B0301030008, the STCSM under Grant No. 23590780100, and the Natural Science Foundation of Shanghai under Grant No. 23JC1400200. Gang Wang is supported by the U.S. Department of Energy under Grant No. DE-FG02-88ER40424 and by the NSFC under Contract No.1835002. Aihong Tang is supported by the US Department of Energy under Grants No. DE-AC02-98CH10886, DE-FG02-89ER40531.

\bibliography{ref}
\end{document}